\documentclass[useAMS, usenatbib]{mn2e}
\usepackage{epsfig}
\usepackage{graphicx}
\usepackage{subfig}

\usepackage[dvips,dvipsnames,usenames]{color}
\usepackage{varioref}
\usepackage{colortbl}
\usepackage[dvips, breaklinks={true}, bookmarks, colorlinks={true}, pdfpagemode={NON}, pdffitwindow=true, pdfstartview={FitB}, linkcolor={BlueViolet}, menucolor={BlueViolet}, citecolor={Mahogany}, filecolor={OliveGreen}, pagecolor={Mahogany}, urlcolor={MidnightBlue}, pdftitle={Calibration SW manual}, pdfauthor={Stephanie Jouvel}, pdfsubject={Calibration}, pdfkeywords={DES,Calibration}]{hyperref}
\usepackage[all]{hypcap}

\usepackage{amsmath,environ}
\bibpunct{(}{)}{;}{a}{}{,}

\title{Optimising Spectroscopic and Photometric Galaxy Surveys: Efficient Target Selection and Survey Strategy} 
\author[S. Jouvel et al.]
{S. Jouvel $^{1,2}$\thanks{E-mail: jouvel@ice.cat}, F. B. Abdalla$^{2}$, D. Kirk$^{2}$, O. Lahav$^{2}$, H. Lin$^{3}$, J. Annis$^{3}$, \newauthor R. Kron$^{3}$, J. A. Frieman$^{3, 4, 5}$. 
\\
\\
$^{1}$Institut de Ciencias de l'Espai, Bellaterra, Spain \\
$^{2}$Department of Physics and Astronomy, University College London, Gower Street, London WC1E6BT, UK \\
$^{3}$Center for Particle Astrophysics, Fermi National Accelerator Laboratory, P.O. Box 500, Batavia, IL 60510, USA\\
$^{4}$Department of Astronomy and Astrophysics, The University of Chicago, 5640 South Ellis Avenue, Chicago, IL 60637, USA  \\
$^{5}$Kavli Institute for Cosmological Physics, The University of Chicago, 5640 South Ellis Avenue Chicago, IL 60637, USA \\} 

\begin{document}
\maketitle

\begin{abstract} The next generation of spectroscopic surveys will have a wealth 
of photometric data available for use in target selection. 
Selecting the best targets is likely to be one of the most important hurdles in making these spectroscopic campaigns as successful as possible
Our ability to measure dark energy depends strongly o
the types of targets that we are able to select with a given photometric data set
We show in this paper that we will be able to successfully select the targets needed 
for the next generation of spectroscopic surveys. We also investigate the 
details of this selection, including optimisation of instrument design and survey strategy in orde
to measure dark energy
We use color-color selection as well as neural networks to select the best possible 
emission line galaxies and luminous red galaxies for a cosmological survey. Using the 
Fisher matrix formalism we forecast the efficiency each target selection scenarios. 
We show how the dark energy figures of merit change in each target selection regime as a function of target type, survey time
survey density and other survey parameters
We outline the optimal target selection scenarios and survey strategy choice
which will be available to the next generation of spectroscopic surveys
\end{abstract}


\date{accepted to MNRAS dec 2013}

\section{Introduction}
\label{sec:intro}

One of the goals of the next decade of cosmological surveys is to 
understand what causes the accelerated expansion of the Universe.
This involves understanding the nature and
behaviour of dark energy and dark matter. To begin to answer these questions, 
several photometric and spectroscopic surveys are being 
planned to start over the next 10 years. 
The Dark Energy Survey\footnote{http://www.darkenergysurvey.org/}, 
hereafter DES, is the first of
a new generation of photometric surveys that focuses on dark matter 
and dark energy studies. 
DES will start observing on the fall of 2013. DES will be followed
by the Euclid \footnote{http://sci.esa.int/euclid/} experiment whose launch is scheduled for 2020 and the Large
Synoptic Survey Telescope (LSST) \footnote{http://www.lsst.org/lsst/} on a similar timescale. On the same
timescale as DES, HypersuprimeCam will map around 2000 square degrees in the 
sky, to greater depth than DES, in similar optical bands.

In terms of spectroscopic surveys, the Sloan Digital Sky Survey (SDSS) and 
BOSS surveys have mapped, or are currently mapping, the large scale structure 
of our Universe. They have targeted Luminous Red Galaxies (LRGs) 
out to redshift $z\approx0.7$.  The next generation of surveys will survey
deeper than SDSS and BOSS by also using the Emission Line Galaxies 
(ELGs). This strategy has already been used by the WiggleZ Dark Energy survey
\citet{Drinkwater10} which targetted ELGs in looking at their NUV flux with 
the Galaxy Evolution Explorer (GALEX) satellite \citep{Martin05}. 
GALEX NUV has a detection limit of 22.8. 
This paper aims to investigate deeper targetting strategy in order to
go up to i$\approx$23.5. We thus study observational
strategies for future spectroscopic surveys using current and future 
photometric surveys such as DES, LSST, Euclid. 

\citet{Abdalla08} showed that we can predict emission lines from 
broadband photometry using Neural network (NN) algorithms. 
In general it would be 
possible to use such techniques in target selection studies. We present
a target selection approach based on the use of NN and broad-band photometry.
Despite the complex selection function that this strategy represents, it can efficiently recover a galaxy's redshift based on spectra
We assess in this paper to what extent several networks can help improve target selectio
and survey strategy

We analyse different strategies of target selection using color-color or more 
complex techniques such as a neural network target selection.
We use different surveys for the photometry depending on the type of galaxies 
we want to select: ELGs at high redshift; LRGs at lower redshift.
Section \ref{sec:survey_definition} describes the goals of future 
cosmological spectroscopic surveys and photometric surveys that can be used 
for target selection. We make our comparisons using mock catalogues 
that we describe in section \ref{sec:mock}. In Section \ref{sec:NNsel}
we outline the neural network target selection we use in this work. In sections
\ref{sec:targetting_lrgs} and \ref{sec:targetting_elgs} we outline our possible
strategies for selection of LRGs and ELGs. Finally, we investigate how the target 
selection affects the survey strategy and the Figures of Merit 
(FoM) for dark energy in section \ref{sec:strategy}.

\section{Current spectroscopic surveys and future plans}
\label{sec:survey_definition}

\subsection{Cosmological spectroscopic surveys}

In order to meet the goals of the current and next generation of cosmological spectroscopi
surveys and improve on our knowledge of dark energy, we nee
 to acquire on the order of a thousend or more galaxy redshift
per square degree up to redshifts of order unity and we need to do this ove
several thousands of square degrees. In this sectio
we outline the possible sources of photometry that can be used to defin
an efficient target selection strategy
Several proposals have been developped such a
DESpec \citep{Abdalla12} which is a spectroscopic survey over the DE
footprint and BigBOSS \citep{Schlegel11}
Both of the above projects have recently been merged into one umbrella project
i.e. the Mid Scale Dark Energy Spectrocopic Instrument (MS-DESI)

Concerning DE experiments, the main interests of spectroscopic surveys is
to help at calibrating systematic errors of photometric surveys such as 
Intrinsic Alignements and photometric redshift uncertainty that affect the WL, cluster counts and BAO
probes. The other main interest is the precise information they provide in the radial direction. 
The large scale photometric surveys will use photometric redshift to set the 
galaxy distances using several photometric bands. The number and width of the bands
is not adapted to high accuracy photometric redshift such as the PAU survey \citep{Gaztanaga12}. 
Spectroscopic redshifts allow us to access radial modes 
at high frequency due to the high precision redshift information.
They are then used to study redshift space distortions (RSDs) which will help tightening DE constraints
when combined with the other DE probes such as shown in the SDSS results of
\citet{Reid12} or the WiggleZ results of \citet{Blake12}. 

BigBOSS was planning to study the Baryon Accoustic Oscillation using
ELGs up to z$\sim$1.7 and quasars up to z$\sim$3 at the NOAO 4‑meter Mayall Telescope 
on Kitt Peak in Arizona. It will survey 14,000 square degrees on the northern hemisphere
and 10,000 square degrees on the southern hemisphere. 
The DESpec survey was planning to cover at least the 
DES footprint of 5000 square degrees. The final strategy for DESI is not yet set on stone.  
The area may extend up to 15,000 square degrees by 
also using LSST photometry for spectroscopic target selection. The DES and LSST 
photometry (together with VISTA JHK photometry) yields not only fluxes and photometric 
redshifts but also galaxy image shapes and surface brightnesses. 
All of this information can be exploited in various ways to select a sample of galaxies 
that satisfy the joint requirements of large redshift range, adequate volume sampling, 
and control over any bias introduced due to sample selection or redshift failures. 
In practice, we expect to use galaxy flux, color (and photo-z), and surface-brightness to 
sculpt the redshift distribution of the survey. 

In this paper, our spectroscopic targetting strategy is to target a mix of emission-line 
galaxies (ELGs)---which predominate at high redshift and which yield efficient 
redshift estimates to z$\sim$1.7 based on their prominent emission lines
---and luminous red galaxies (LRGs), with brighter continuum spectra, 
which offer high redshift success rates at lower redshifts, up to z$\sim$1. 
The target selection will ultimately be chosen to optimize the science yield; 
at this stage, we present some examples of survey selection to demonstrate 
its feasibility. 

\subsection{Current and future surveys useful to target selection}

The photometry from current and future surveys will be useful to
the selection of targets in DESI and other spectroscopic surveys
such as eBOSS \footnote{http://www.sdss3.org/future/eboss.php}, 
SUMIRE\footnote{http://sumire.ipmu.jp/en/}, 4MOST \citep{4MOST}. 
In order to test the
different target selection strategies and their impact on DE
science outputs, we simulated different surveys with mock catalogues that 
we describe in section \ref{sec:mock}.

We aim at covering a wide area on the sky with a wavelength range that will allow the
study of galaxy population and large scale structure at high redshift.  
Past and current spectroscopic surveys are either large and relatively shallow such as
BOSS and SDSS\footnote{http://www.sdss.org/}, which cover several thousands of square degrees 
up to z $\approx$0.7, or deep on a small field such as Wigglez\footnote{http://wigglez.swin.edu.au/site/}, 
VVDS\footnote{http://cesam.oamp.fr/vvdsproject/},
zCOSMOS\footnote{http://www.exp-astro.phys.ethz.ch/zCOSMOS/}, and VIPERS\footnote{http://vipers.inaf.it/}.
These deep surveys can go up to z$\approx$6 but rarely cover more than several hundreds of square degrees.
We need to define a target selection based on large photometric surveys.
To compare the impact of photometric depth in the target selection, we simulate the photometry of
optical surveys such as the DES,
the Palormar Transient Factory (PTF)\footnote{http://www.astro.caltech.edu/ptf/},
and PanSTARRS (PS1)\footnote{http://pan-starrs.ifa.hawaii.edu/public/}.
Between them these surveys represent the range of area/depth options currently under consideration. 
The photometry from these
surveys allows for different target selections that form the basis of this paper.

To complement optical surveys, the DES footprint overlaps with the
VISTA/VHS\footnote{http://www.ast.cam.ac.uk/~rgm/vhs/}. This produces better 
photometric redshifts as studied in \citet{Banerji08} and
will help the target selection for galaxies at $z>1$. In the next sections,
we will consider a target selection based on deep photometry using the
DES+VISTA surveys.

We also simulated WISE\footnote{http://wise.ssl.berkeley.edu/} photometry.
Galaxies show a 1.6$\mu$ m bump restframe that will be used in the target selection
of higher redshift galaxies with the w1 photometry. We detail the different
target selection in the next sections depending on the type of galaxies we are
targetting such as LRGs in section \ref{sec:targetting_lrgs} and ELGs in
section \ref{sec:targetting_elgs}.

\begin{figure}
\resizebox{\hsize}{!}{\includegraphics{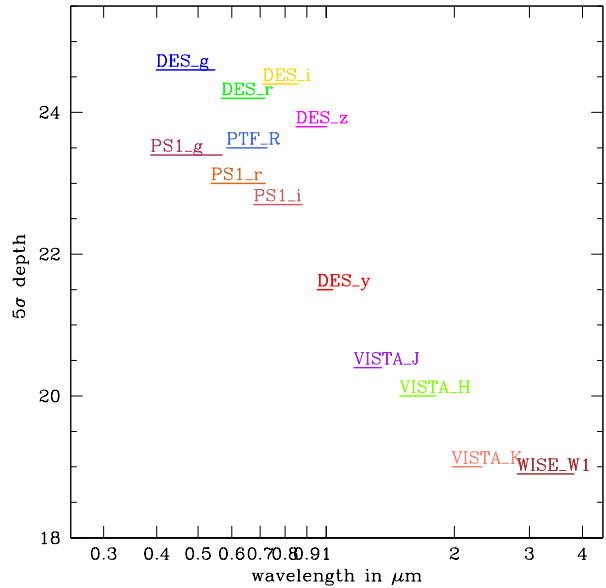}}
 \caption{Band width in $\mu$m and 5$\sigma$ depth of extended sources for the DES survey, PanSTARRS 1 year survey, PTF R band and
VISTA JHK bands. }
\label{fig:surveys_filters}
\end{figure}

To give a quick overview of these surveys:
\begin{itemize}
\item DES will cover 5000 deg$^2$ in 5 photometric bands covering 0.4 to 1 $\mu$m
to a depth of i-band 24 mag AB at 5$\sigma$ in the aim of studying the
large scale structures from the cosmic shear and BAO signal, the supernovae and
galaxy clusters. The first observations will start in november 2012 and last for
5 years. DES fields are located in the southern hemisphere and are visible
during 4 months at the end of the year.
\item PTF is an ongoing transient survey targeting 8000 deg$^2$/yr
in the 2 photometric bands R and g' to study mainly supernovae, binaries and
transiting planets. The survey will last 5 years and will do a full sky area.
\item panSTARRS is an ongoing survey that will cover a full sky area in 5 photometric
bands covering the same wavelength range as the DES survey. panSTARRS strategy is
divided in 4 stages. Each stage will complete a full sky and increase the
deth from the precedent stage. In this paper, we simulate the photometry of stage 1.
\item VISTA/VHS will cover 19000 deg$^2$ in the southern hemisphere to study the Milky Way,
discover nearby stars and low mass stars, study Dark Energy out to z$\approx$1 and high
redshift quasars using NIR photometry.
\item WISE will survey the entire sky in mid-IR photometry to study asteroids of
our solar systems, nearby stars, the Milky Way, and AGN.
\end{itemize}
We give the depth at 5$\sigma$ of the different photometric bands of the
surveys that we describe in this section in Table  \ref{tab:survey_depth}.
We concentrate on the bands that we are planning to use in the target
selection. We show the bands wavelength coverage and 5 $\sigma$ depth for
extended objects in \autoref{fig:surveys_filters}.
\begin{table}
  \caption{Photometry of surveys used in for the targets selection. The table shows the depth of extended sources at 5$\sigma$.}
\begin{tabular}{cccccccc} \hline\hline Survey & Band & AB mag(5$\sigma$) \\
  DES & $g$    &   24.6    &   \\
  DES & $r$    &   24.2    &   \\
  DES & $i$    &   24.4    &   \\
  DES & $z$    &   23.8    &   \\
  DES & $y$    &   21.5    &   \\
  VISTA & $J$  &   20.4    &   \\
  VISTA & $H$  &   20.0    &   \\
  VISTA & $K$  &   19.0    &   \\
  PS1   & $g$  &   23.4    &   \\
  PS1   & $r$  &   23.0    &   \\
  PS1   & $i$  &   22.7    &   \\
  PTF   & $R$  &   23.5    &   \\
  WISE   & 3.6um  &   18.9    &   \\
 \hline \hline
\label{tab:survey_depth}
\end{tabular}
\end{table}

\section{Spectro-photometric simulations}
\label{sec:mock}

\subsection{The COSMOS Mock Catalogue}
To quantify the yield of a particular spectroscopic scheme, we use the COSMOS Mock Catalogue (CMC)
\citet{Jouvel09,Zoubian13}.
The CMC is built from the COSMOS photometric-redshift catalogue \citet{Capak08,Ilbert09}. 
We use the CMC to predict the target number density and to compute the spectroscopic instrument
sensitivities. 
The COSMOS photometric-redshift catalog \citep{Ilbert09} was computed with 30 bands from GALEX for the UV bands,
Subaru for the optical, CFHT, UKIRT and Spitzer for the NIR bands over 2 deg$^2$. 
The CMC is restricted to the area fully covered by
HST/ACS imaging, 1.24 deg2 after removal of masked areas. There are a total of 538,000 simulated
galaxies for $i<26.5$ leading to a density of roughly 120 gal/arcmin$^2$.
The CMC includes AGN and star.
COSMOS have high quality photo-z with an excellent accuracy and low catastrophic redshift rates
from a careful calibration with the spectroscopic samples zCOSMOS and MIPS. 
To derive properties used in the simulations from each observed COSMOS galaxy, a photo-z and a 
best-fit template spectrum (including possible additional extinction) are associated with each galaxy of 
the COSMOS catalog. We first integrate the best-fit template through the instrument filter 
transmission curves to produce simulated magnitudes in the instrument filter set. 
We then apply random errors to the simulated magnitudes based on a simple magnitude-error 
relation in each filter. The simulated mix of galaxy populations is then, by construction, 
representative of a real galaxy survey, and additional quantities measured in COSMOS (such as 
galaxy size, UV luminosity, morphology, stellar masses, correlation in position) can be easily 
propagated to the simulated catalog. The COSMOS mock catalog is limited to the range of magnitude 
where the COSMOS imaging is complete ($i_{AB} \approx 26.2$ for a 5 $\sigma$ detection \citep{Capak07,Capak08}. 
We associate emission-line fluxes for each galaxy of the CMC. We model the emission-line 
fluxes of each galaxy using the \citet{Kennicutt98} calibration which links the 
star-formation rate (SFR) from the dust-corrected UV rest-frame luminosity already measured 
for each COSMOS galaxy. The SFR is then translated to an 
[OII] emission-line flux using another calibration from \citet{Kennicutt98}. We model the Balmer 
lines (Ly, Hb, Hd, Hg, Ha), the oxygen lines (OII, OIIIa, OIIIb), the nitrogen line (NII) 
and the sulfur lines (SIIa, SIIb) from emission line ratios \citet{McCall85,Moustakas06,Mouhcine05,Kennicutt98}
and zCOSMOS spectroscopic data.
The relation found 
between the [OII] fluxes and the UV luminosity is in good agreement with the VVDS data and still 
do not vary for different galaxy populations as shown in \citet{Jouvel09}. 
\citet{Jouvel09} validates the CMC and show that it reproduces the counts and color
distributions from the optical to NIR bands in comparing to the GOODS \citep{Giavalisco04} and
UDF \citep{Coe06} surveys. The CMC also provides an excellent match to the redshift-magnitude 
and redshift-color distributions for I$<$24 galaxies in the VVDS spectroscopic redshift survey \citep{LeFevre05}. 

\subsection{Sensitivities and signal-to-noise for spectroscopic surveys}
\label{sec:sensitivities}

We choose here one spectrograph configuration which corresponds to a DESpec-like survey but
stress that the optimisation procedure in further section remain applicable to any target selection study.
We simulate the spectroscopic instrument throughput, sensitivities, exposure time, and success rate using
the simulated galaxy spectra produced by the COSMOS Mock Catalogue (CMC, \citet{Jouvel09}) and from
the DES galaxy catalogue simulations (R. Wechsler \& M. Busha). Both model galaxy physics such
as emission lines, dust extinction, at redshift ranges and magnitudes appropriate to our
spectroscopic survey.

The DES catalog simulations provide galaxy spectra based on the template models 
generated by the Kcorrect package \citet{Blanton07}, while as just described 
the CMC provides spectra based on the COSMOS data, providing in particular detailed 
distributions of line fluxes for emission-line galaxies. We then fold in the various 
contributions to the overall system throughput, speciﬁcally the transmission 
vs. wavelength for the atmosphere, the telescope and spectrograph optics, and the fibers, 
plus the quantum efficiency of the CCD detectors. 

In this study we have assumed a 50\% throughput efficiency 
for the spectrograph plus fibers and used the size distribution of our galaxies
to obtain the correct amount of light gathered. We have used standard cross correlation
techniques to obtain the spectroscopic efficiency of our 
instrument once noise has been added to the simulations. We
have assumed 2-arcsec-diameter fibers and a resolution of 
$R=3500$ at 900nm for these calculations.  
For the detailed calculations for the sensititvies of the 
fidutial spectrograph we refere the reader to \citep{Abdalla12}.

\section{Target selection: NN and color-color}
\label{sec:NNsel}

Several surveys used color-color selection to sculpt the redshift
distribution in order to study specific redshift ranges of the galaxy population 
such as DEEP2, VIPERS. DEEP2 select
galaxies between 0.75$<$z$<$1.4 with photometry in BRI bands, VIPERS select galaxies from 
0.5$<$z$<$1.4 with photometry u\*gri bands. The resulting surveys show that it is possible
to select galaxy targets based on color-color selections which work well 
to realise a broad redshift selection. 

In the future, during the next five years, the DES and VISTA surveys in the south, and in the next ten years
LSST and Euclid over the whole sky, will provide rich photometric data on wide wavelength
ranges from optical to NIR. In this paper, we present a way of utilising DES and VISTA broadband photometry
in order to do an optimized galaxy target selection. We show a new technique based on 
NN fed with the photometry from surveys such as those mentioned above. 
A NN technique is similar to a multi-dimensional color-color selection. It
assumes multi-color informations for 4 or more bands. 
The network captures emission line features from ELG colors using the extra color 
information provided by the multi-band surveys. 

Throughough the paper, we use ANNz \citep{Collister04}.
\citet{Abdalla08nn} shows that one can predict 
emission line strength from galaxy colors. As we briefly explained in section
\ref{sec:mock}, the CMC simulation predicts emission lines fluxes and include them 
in the galaxy magnitudes derived. Using the CMC multi-color
information from the DES+VHS surveys, we teach the neural network to recognise 
ELGs using a representative training sample of 10000 galaxies. 

\begin{figure}
\resizebox{\hsize}{!}{\includegraphics{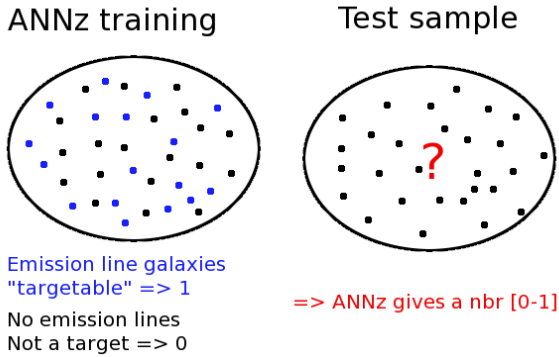}}
  \caption{Schematic representation of the targetting strategy using Neural Network.
The blue dotted points correspond to the galaxies with strong emission lines for which we will be able
to measure a redshift using our spectroscopic sensitivities. The black dot correspond to 
the galaxies with weak or none emission lines galaxies.}
\label{fig:nn_target_schem}
\end{figure}

\autoref{fig:nn_target_schem} is a schematic representation of the NN target selection. 
The left bubble represents a training sample for which the
blue dots symbolise strong emission line galaxies that we want to target, and the black dots
other types of galaxies. To realise the selection, we assume a telescope model using the 
sensitivity curves described in section \ref{sec:sensitivities}.
We train the NN using a target/not-a-target selection and test it on a sample
equally representative of the training and validation sample.  
The NN outputs a number which can be associated with 
a probability of the galaxy being of the type you trained the NN to recognise. 
We then trade off between galaxy number density and purity of the sample in 
choosing a probability value that defines whether the galaxy is a target. 

We use the NN target selection to select ELGs and LRGs following the procedure
we describe in this section.
We also realise 2 more traditional color-color selections to compare
results. We detail below the 3 target selection methods we study in this paper.

\begin{itemize}
\item Shallow color-color target selection: photometry from panSTARRS, PTF and WISE
\item Deep color-color target selection: photometry from DES+VHS
\item NN target selection: photometry from DES+VHS
\end{itemize}

We further compare the different targetting strategies for LRGs and ELGs in section 
\ref{sec:targetting_lrgs} and section \ref{sec:targetting_elgs}.

\section{Targetting strategy for LRGs}
\label{sec:targetting_lrgs}

\subsection{Strategy and plans for LRGs}
\begin{figure}
\resizebox{\hsize}{!}{\includegraphics{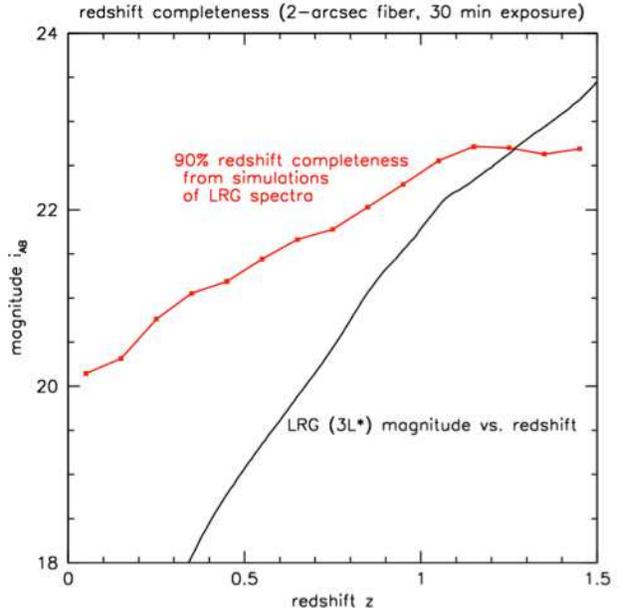}}
\caption{The red points and curve indicate the $i_{AB}$ magnitude vs. redshift where the redshift
success rate for an LRG spectrum is 90\%, based on our spectroscopic simulations.
Also plotted in black is the magnitude vs. redshift relation for an LRG, taken as
a 3L* galaxy (Eisenstein et al. 2001), showing that we will be 90\% successful
for LRG redshift measurements out to $z=1.3$ in a 30 minute exposure.}
\label{fig:completness_lrg}
\end{figure}

In order to measure the BAO signal to redshift z$\approx$1 we need of order $\approx$300 
luminous red galaxies per sq. deg. Results from SDSS and BOSS indicate that LRG targets 
in this redshift range should have a high rate of redshift success. For galaxy clustering 
measurements, LRGs are convenient because they occupy dense regions and are thus strongly 
biased relative to the dark matter, b$\approx$2 as quoted in \citet{Dawson13}. With our simulations we have estimated the 
completeness rate for objects in the CMC catalogue. We also plot in  
\autoref{fig:completness_lrg} a preliminary 
example of a redshift completeness for LRGs, showing the magnitude, as a function of redshift, 
where the redshift success rate is expected to be 90\%, based on the above simulations 
and assumptions. Also shown on the plot is the apparent magnitude vs. redshift relation 
for an LRG, taken to be a 3L* galaxy \citep{Eisenstein01}, showing that we will be 90\% 
complete for LRG redshift measurement out to z = 1.3 in a 30 minute exposure. Given that 
the redshifts of these galaxies are attainable by our spectroscopic instrument out to redshifts just higher than 
one with a 30 min exposure time we aim to obtain a target selection from the photometric data.

\subsection{LRGs target selection using deep optical-NIR photometry}
\begin{figure}
\resizebox{\hsize}{!}{\includegraphics{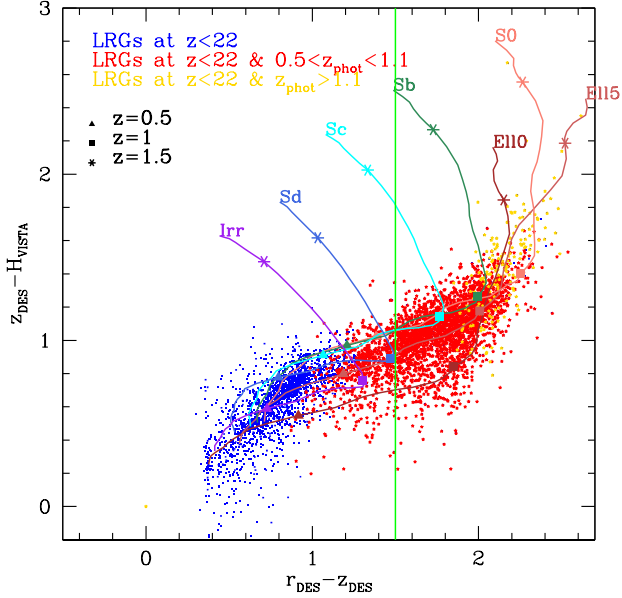}}
 \caption{LRGs target selection using DES+VHS photometry. Example of the targeting efficiency
 of color-color cuts for galaxies at magnitude z$<$22, in blue for a 
redshift z$<$0.5, green for 0.5$<$z$<$1.1, and gold for z$>$1.1.}
\label{fig:des_lrg_sel}
\end{figure}

Using DES+VHS imaging, we can target the LRGs with z-mag less than the depth of DES and yielding a relative flat 
redshift distribution over the range 0.5$<$z$<$1. We use selection cuts in r-z vs. z-H color space
in addition to a magnitude selection of $z_{DES}$. 
We show the colour diagrams in \autoref{fig:des_lrg_sel}.
These cuts supply more than the needed density 
of LRG targets, so they can be randomly sampled. This r-z cut  

\subsection{Target selection using shallow optical-IR photometry}
\begin{figure}
\resizebox{\hsize}{!}{\includegraphics{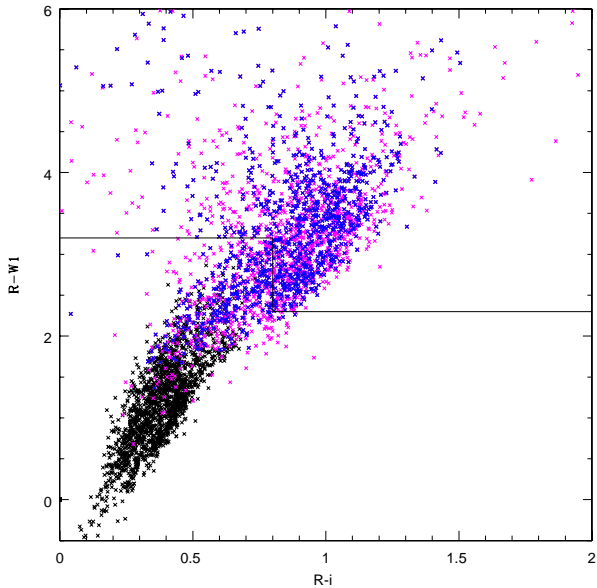}}
 \caption{LRGs target selection using PTF+WISE photometry. This figure shows 
R-i as a function of W1-R colors in black for all galaxies for which W1$<$18.9.
We also show galaxies at z$>$0.55 for which  W1$<$18.9 in magenta and W1$<$19.9
in blue.}
\label{fig:lrg_bboss}
\end{figure}

As a comparison, we use the CMC simulation of the PTF, panSTARRS and WISE photometry and
reproduce the target selection of the BigBOSS surveys. \autoref{fig:lrg_bboss}
shows R-i from the PTF and panSTARRS surveys versus R-W1 color from the PTF and WISE surveys.
The LRGs selection consist in targetting all the galaxies that are above
the black line. In black we show all galaxies for which W1$<$18.9.
In magenta and blue we show galaxies at $z>0.55$ for which respectively W1$<$18.9 and W1$<$19.9.
The PTF+WISE selection is aiming to select bright LRGs based on the 1.6um bump 
for galaxies at $z>0.5$. This selection allow to select a sample of galaxies
whose mean redshift will be around $z\approx1$. 
 
\subsection{LRGs target selection using NN}
\begin{figure}
\resizebox{\hsize}{!}{\includegraphics{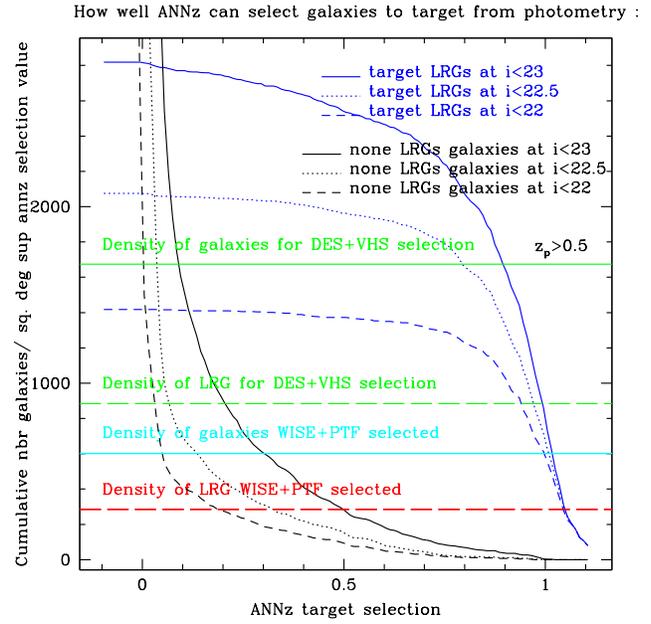}}
  \caption{Cumulative number of galaxies as fct of ANNz target selection for LRG.
Overplotted the number density for the color-color selections}
\label{fig:cumul_lrg}
\end{figure}

We use DES and VHS photometry and train the NN to select
elliptical galaxies at i brighter than 23 AB magnitude without redshift
restriction. The CMC is built from the best-fit
template and photometric redshift of the COSMOS survey. 
Using the COSMOS galaxy types of the CMC we are able to select
elliptical galaxies from DES+VHS colors using NN such as explained in
section \ref{sec:NNsel}.  
\autoref{fig:cumul_lrg} shows the cumulative distribution of ANNz
probability distribution values for galaxies brighter than i $\approx$ 
23, 23.5, and 24 in respectively small-dashed, dotted and solid lines. The blue
lines are the cumulative number of LRGs selected by the NN while the black lines are
other types of galaxies. The green lines give the number density of galaxies
selected using the color-color cut with DES+VHS photometry in solid lines and
the density of LRGs inside the color-color cut in long-dashed lines. In a same 
way, the cyan lines represent the number density of galaxies and LRGs selected
using the PTF+WISE photometry cut.  

\clearpage 
\subsection{Redshift distribution of color and NN selection for LRGs}
\begin{figure}
\resizebox{\hsize}{!}{\includegraphics{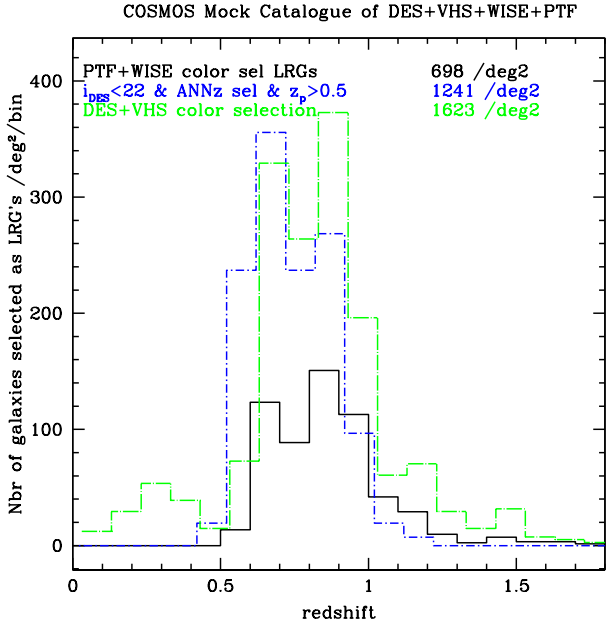}}
 \caption{Redshift distribution for the color-color selection PTF+WISE and DES+VHS 
and the NN selection in respectively black solid, long dashed-dotted green and small
dashed-dotted blue lines. We realise a cut in the NN selection at $i_{DES}<22$ with
a photometric redshift at $z_p>0.5$.}
\label{fig:nz_lrg}
\end{figure}

Here we compare the redshift distribution of the different targetting strategies.
\autoref{fig:nz_lrg} shows the redshift distribution of the 2 color-color section PTF+WISE,
DES+VHS and the NN selection in respectively black solid, long dashed-dotted green and small
dashed-dotted blue lines. For the NN selection, we add a photometric redshift cut at
$z_p>0.5$ since we are interested mostly in the high-redshift LRGs at $0.5<z<1$ as explained
in the survey definition section \ref{sec:survey_definition}. 

\begin{figure}
\resizebox{\hsize}{!}{\includegraphics{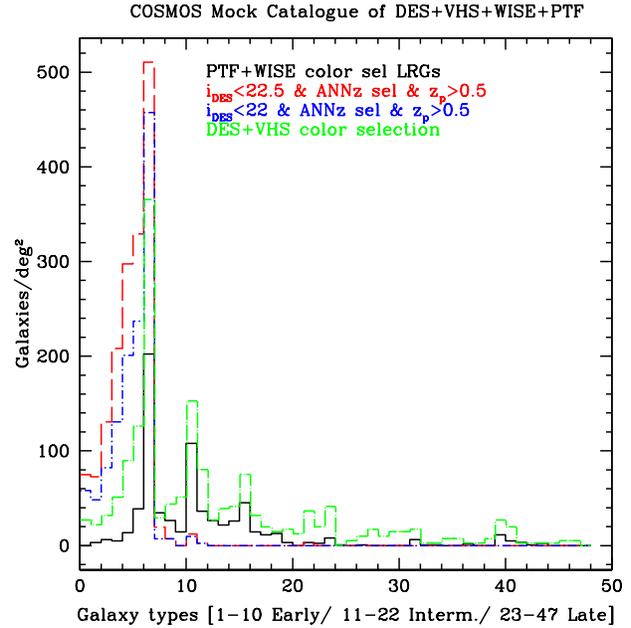}}
 \caption{Histogram of types for the color-color PTF+WISE and DES+VHS and NN selection
strategy in respectively black solid, long dashed-dotted green and small
dashed-dotted blue lines. We realise a cut in the NN selection at $i_{DES}<22$ with
a photometric redshift at $z_p>0.5$.}
\label{fig:typehist_lrg}
\end{figure}

The DES+VHS color-color selection yields the highest number of LRGs with a density of 
1623gal/deg$^2$ compared to 1241gal/deg$^2$ for the NN selection and 698gal/deg$^2$ for
the PTF+WISE selection. However the color-color targetting strategy has a broader range
of galaxy types as you can see in \autoref{fig:typehist_lrg}. 
\autoref{fig:typehist_lrg} shows the distribution of galaxy types
for each of the targetting strategy: color-color using DES+VHS in green 
long dashed-dotted lines, color-color using PTF+WISE in black solid line, the NN and
photometric redshift selection in blue small dashed-dotted lines at $i_{DES}<22$ and
in red dashed lines for $i_{DES}<22.5$. 
The types corresponds to the library templates
used to derive the photometric redshift of the COSMOS survey. 
There is 8 ellipticals, 11 spirals and 12 starbursts from \citet{Bruzual03} and \citet{Polletta07}
calibrated using the zCOSMOS and MIPS spectroscopic redshift surveys as explained in section
\ref{sec:mock}. In this figure, the early type galaxies ranges from 1 to 10, followed by 
intermediate types from 11 to 22 and late types from 23 to 31.
This figure shows that both color-color selection targets some intermediate
and late types galaxies in their LRGs sample. The LRG targetting strategy relies
on a redshift measurement based on the spectrum of the continuum. The intermediate and
late types are less reliably determined from their continuum. This will likely
lower the efficiency of the target selection. 
The PTF+WISE CC selection is shallower and have less contamination from
the intermediate and late types. However in proportion, it is as affected as the
DES+VHS CC selection.
The NN selection works well at selecting the early-types galaxies from DES+VHS
photometry and is not contaminated by intermediate and late types. The
spectroscopic success rate will be higher in a NN selection than a CC selection.
However, an NN+photoz selection is more complexe than a CC selection 
in terms of selection function. For clustering and galaxy evolution studies,
you need to determine the completness of the galaxy sample. 
A NN+photoz selection will give more work on the understanding of the selection function.

\subsection{Spectroscopic success rates of LRG target selection}
\begin{figure}
\resizebox{\hsize}{!}{\includegraphics{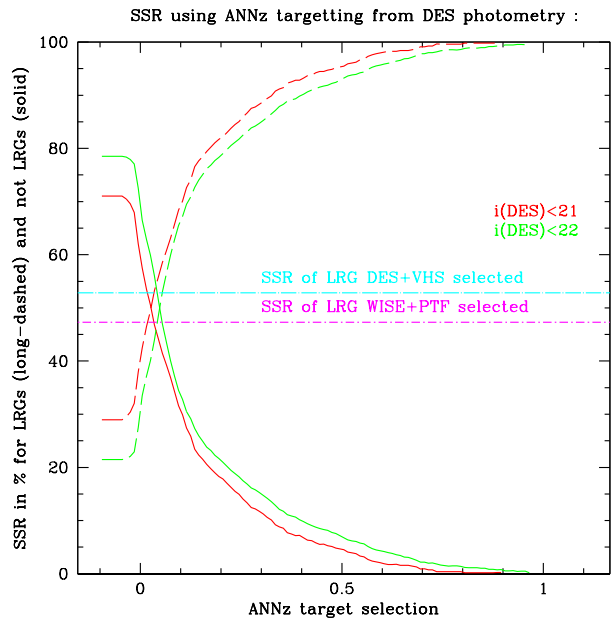}}
 \caption{Spectroscopic Success Rate as a function of ANN selection values for
the LRGs selected using DES+VHS photometry in long-dashed lines and for other types
of galaxies in solid lines. The red and green lines select galaxies at respectively i 
brighter than 21 and 22. The long-dashed and small-dashed dotted lines are respectively
the success rate of DES+VHS and PTF+WISE LRG color selection.} 
\label{fig:ssr_lrg}
\end{figure}
\autoref{fig:ssr_lrg} show the Spectroscopic Success Rate of the
LRGs selection with ANNz in long-dashed lines at i brighter than 21 and 22
in respectively red and green lines. The other types of galaxies are shown
in solid lines. A selection cut at 0.8 gives very high success
rate with a NN selection. For i brighter than 21, there is no contamination
by other types of galaxies and a very small percentage for i brighter than 22. 
As a comparison, we show the PTF+WISE and DES+VHS color selection in respectively
small-dashed dotted magenta line and long-dashed dotted cyan line. 
Both color selections show similar success rates with respectively 
48\% and 53\% for WISE+PTF and DES+VHS color selection of LRGs. 
Color selections
allow a secure redshift measurement for about half of the galaxies targetted while
the NN selection allows close to 98\% redshift measurement. However, this study relies
on several strong assumptions such as: 1. a perfect knowledge of galaxy populations 
2. a non-biased photometry in the photon noise regime 3. a representative
training sample. In a real survey, to perform this strategy, a few nights of 
spectroscopic data would be needed to train the NN response to this selection. 

\autoref{fig:nz_lrg} show that the NN selection reaches a number of galaxies close
to the one reached using a color selection with the same photometry, DES+VHS
in this figure. The NN gives a success rate close to 100\%. However, we know
that all the assumptions we made are unrealistic which will lower down 
the success rate of the NN and possibly the color selection as well.  
We will need to test both methods on more developped simulations of the instrument
and on real data for a definite validation.

\section{Targetting stragety for ELGs}
\label{sec:targetting_elgs}

In this section, we describe simulations to investigate selection criteria 
for a survey targeting emission-line
galaxies at redshifts higher than accessible with luminous red galaxies
for BAO and RSD studies. Using the CMC \citep{Jouvel09}, we assume that the 
survey would reach the line sensitivities described earlier, namely a signal-to-noise 
ratio of around 5 for fluxes larger than 3e-17 erg/s/cm2 from wavelengths 600 to 1000 nm. 
This is reached for a 30 min exposure, which we take as the baseline. 

We compare color target selection and NN selection for ELGs at
$z>1$ in the same way we did in section \ref{sec:targetting_lrgs}.

\subsection{Target selection using color cuts in the optical}

DEEP2, VIPERS, Wigglez and other surveys used color selection
in order to select emission line galaxies at some redshit ranges. 
The BigBOSS survey is planning to use PTF and panSTARRS photometry 
to realise the target selection. In this section, we reproduce the
PTF-panSTARRS target selection using mock photometry of these surveys 
simulated using the CMC, see section \ref{sec:mock}.

\begin{figure}
\resizebox{\hsize}{!}{\includegraphics{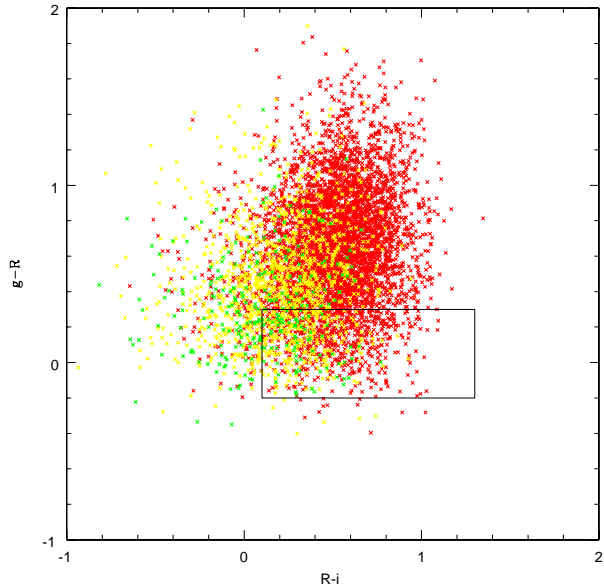}}
 \caption{Color-color selection of ELG galaxies. The x-axis shows R-i
of panSTARRS i band and PTF R band as a function of g-R of panSTARRS g band
and PTF R band. The black box represent the color-color cut we apply
to selec bright ELG galaxies at 0.7$<$z$<$2.}
\label{fig:elg_col}
\end{figure}
\autoref{fig:elg_col} show  R-i
of panSTARRS i and PTF R band as a function of g-R of panSTARRS g 
and PTF R band. The black box represents the color cuts we apply
to select bright ELG galaxies at 0.7$<$z$<$1.7. The green, yellow, and red
points are galaxies at respectively 0.7$<$z$<$1.2, 1.2$<$z$<$1.6, and 1.6$<$z$<$2.

\subsection{Target selection using NN techniques}

\begin{figure}
\includegraphics[width=\columnwidth,height=0.9\columnwidth]{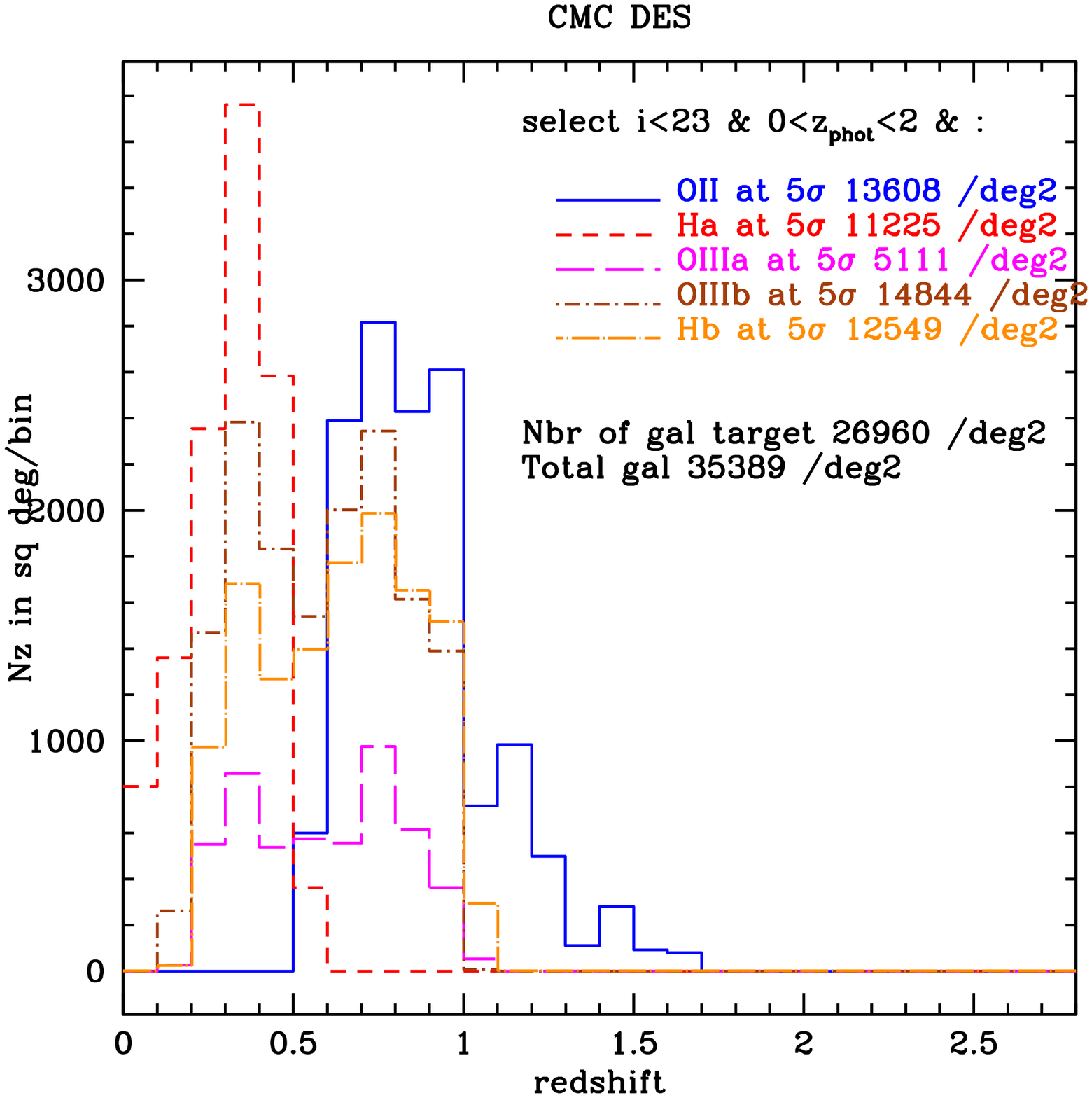}
\includegraphics[width=\columnwidth,height=0.9\columnwidth]{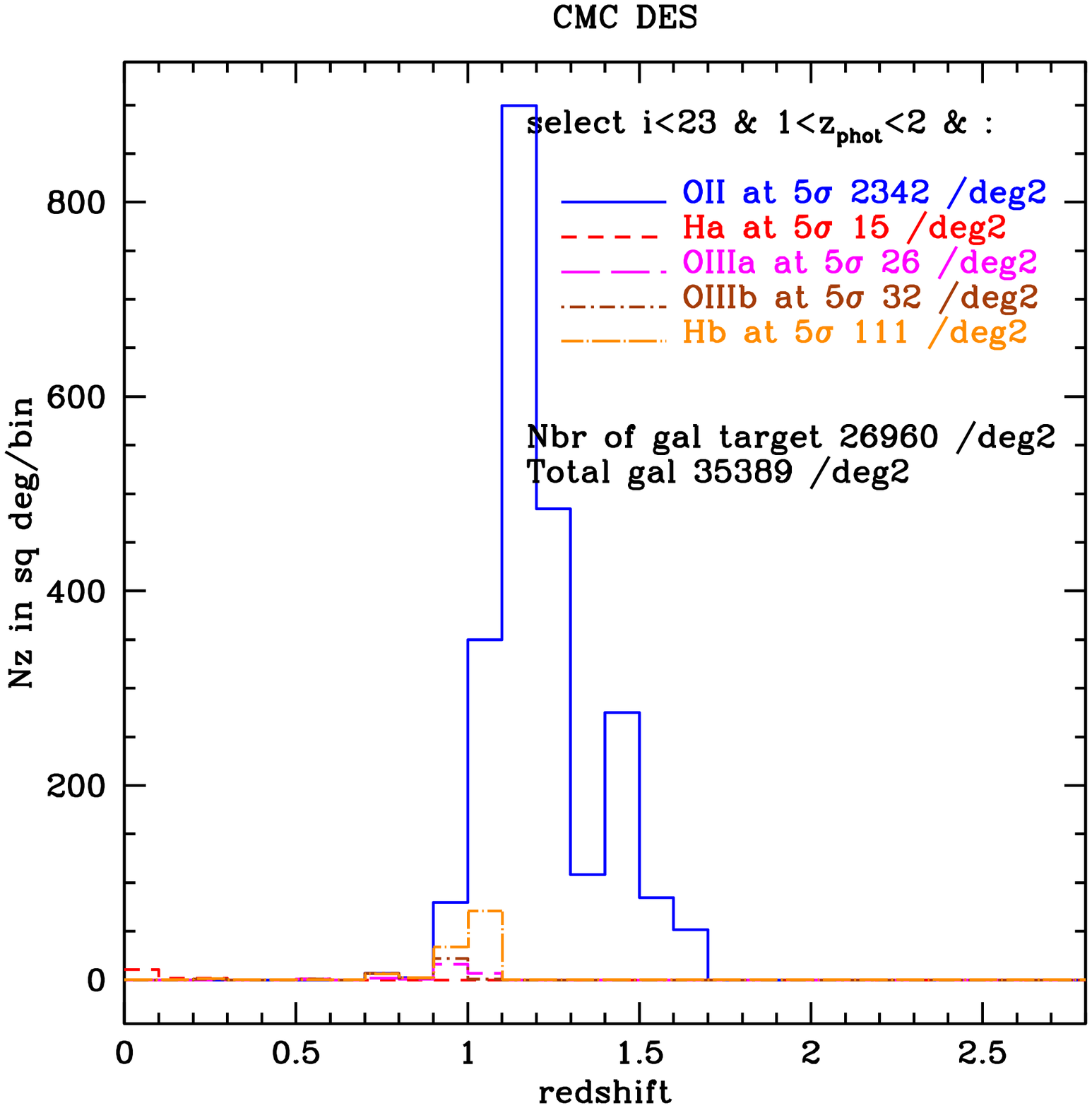}
\includegraphics[width=\columnwidth,height=0.9\columnwidth]{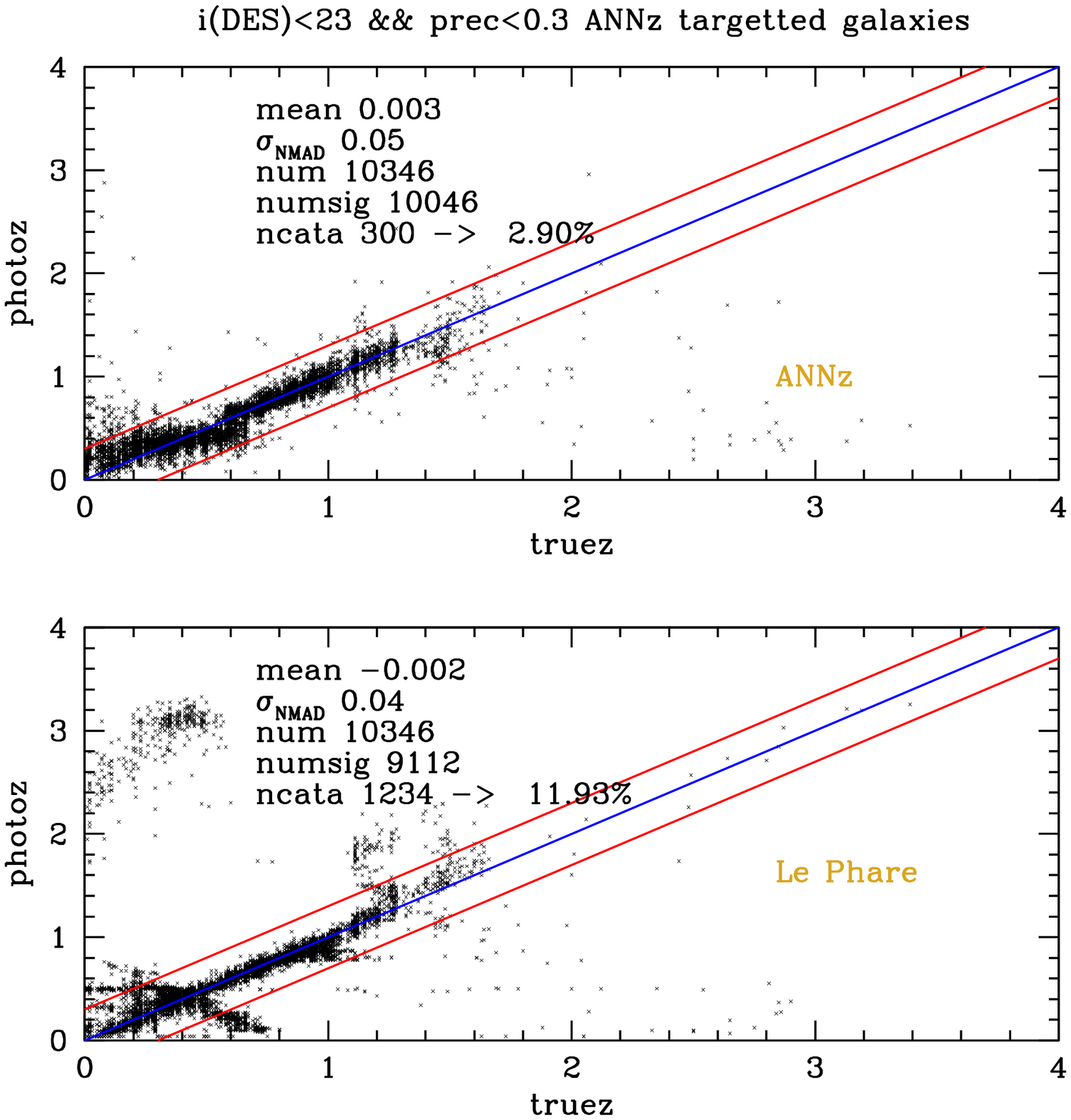}
\caption{Example of the yield of a particular target-selection scheme based on the COSMOS Mock
Catalogue (CMC) simulations.  Here galaxies brighter than i=23 are selected that have photometric
redshifts between 1 and 2 (left) and the full available number of targets is plotted in the middle panel,
with photometric redshifts between 0 and 3. The left panel compares the photoz of ANNz and Le Phare.}
\label{fig:photoz_sel} 
\end{figure}

We can use photometric redshifts to select for high-redshift galaxies. 
\autoref{fig:photoz_sel} shows the redshift distribution of emission-line galaxies 
that our spectroscopic instrument would be able to target 
with a photometric redshift cut between $1<z_{phot}<2$. (This color-color selection, 
which in fact will be similar to the one that was done for the DEEP2 redshift survey of 
ELGs over the redshift range 0.7-1.4, gives a similar result but is instead based on the 
photo-z calculation alone, especially for template-fitting code whose results are based 
on a chi-squared fitting procedure.) \autoref{fig:photoz_sel} also shows the distribution 
from a redshift 
measurement based on a S/N=5 detection of the OII line in blue, H$\alpha$ in red, OIIIa in magenta, 
OIIIb in brown, H$\beta$ in gold. Exercises such as this give a first indication of the spectroscopic 
success rates, redshift distribution, and numbers of galaxies that would result from such a survey. 
We have assumed a 5 sigma detection limit from the OII as well as other lines. 
A wavelength range 600$-$1000 nm allows H$\alpha$ to be detected up to z=0.52 and [OII] to be detected 
at z$>$0.6; the upper limit in redshift for [OII] is z=1.7. We reach a total number of 2500 
galaxies/deg2 between redshift 1 and 1.7 for galaxies with i$<$23. Additional redshift measurements 
will come from lines other than [OII], totaling 300 gal/deg2. 

\autoref{fig:photoz_sel} shows that selection of high-redshift sources via photometric redshift is effective, and it shows 
that, as expected, [OII] is the principal spectroscopic feature in this range of redshifts. 
Furthermore all redshifts can be accessible with other lines including H$\alpha$, H$\beta$ and OIII. 
In \autoref{fig:photoz_sel} we show the photometric redshifts plots for the same sample of galaxies. 
This should allow us to estimate the redshifts we miss due to the photo-z selection. For galaxies 
selected between redshifts 1 and 2 4.5\% of galaxies are missed by this selection. i.e. 4.5\% of 
galaxies will be outside the desired redshifts range. If we choose instead to select a low redshift 
emission line sample, the blunder rate with photo-z selection will be 12\% of the galaxies.

\begin{figure}
\includegraphics[width=\columnwidth,height=0.9\columnwidth]{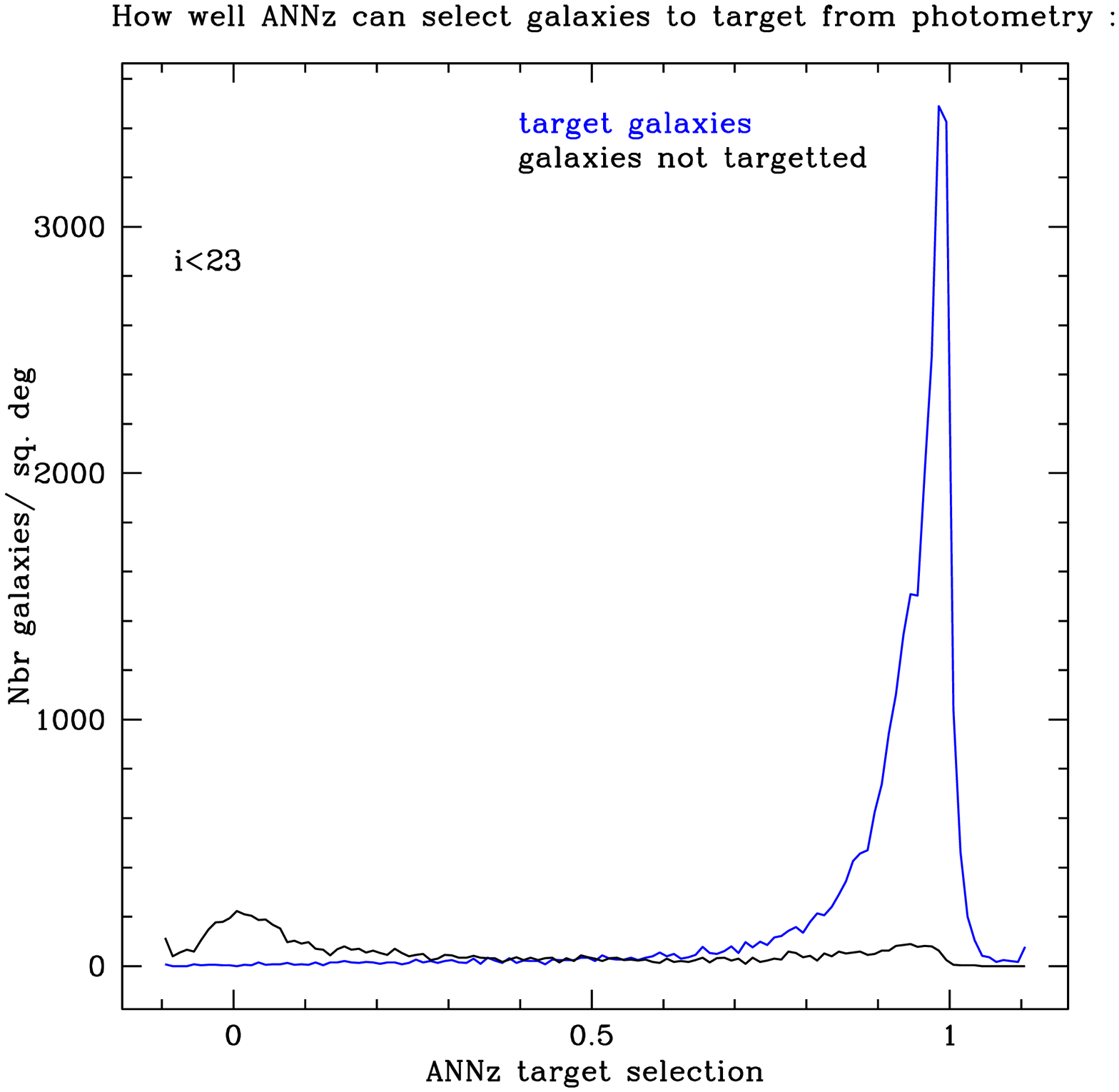}
\includegraphics[width=\columnwidth,height=0.9\columnwidth]{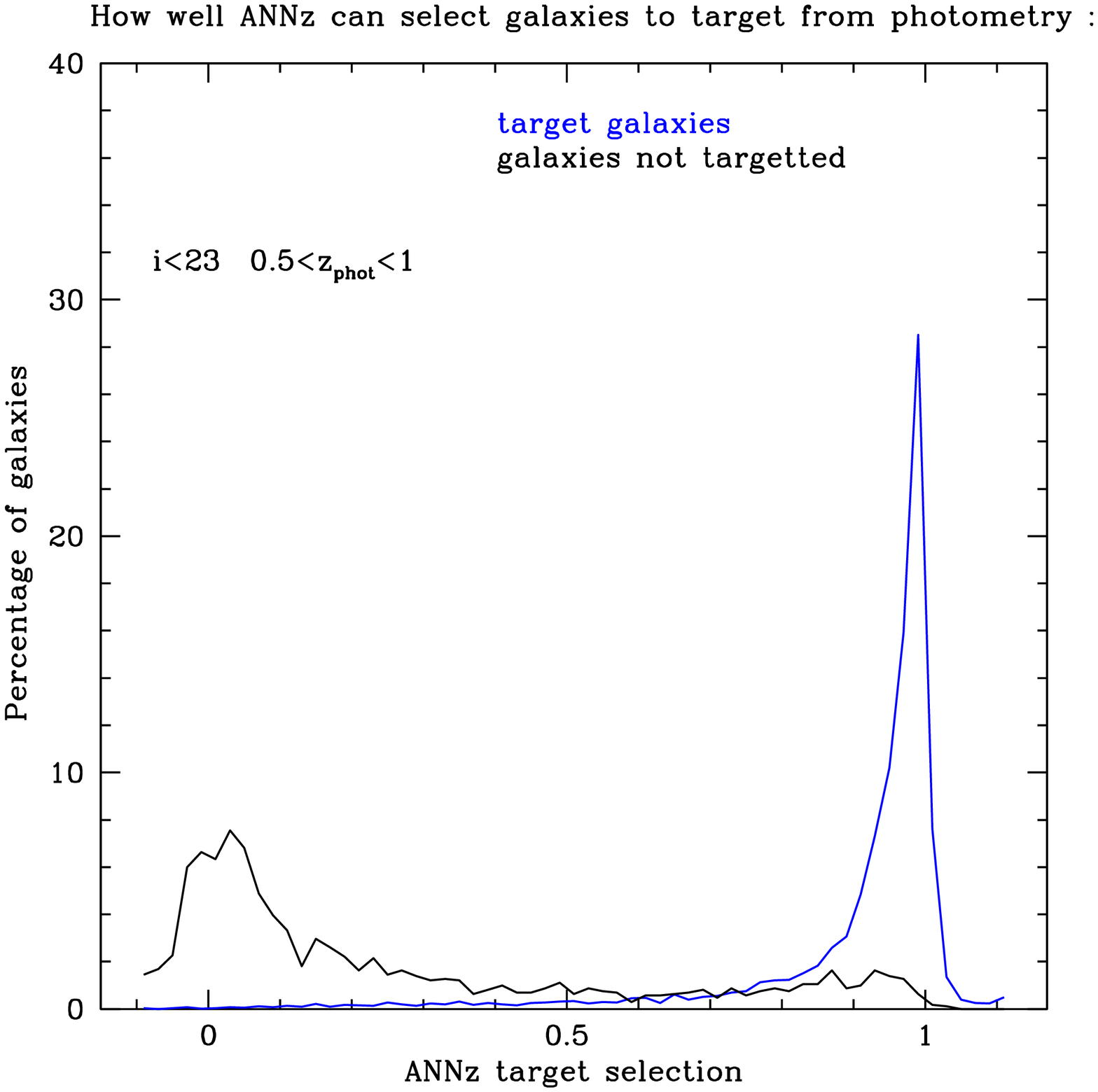}
\includegraphics[width=\columnwidth,height=0.9\columnwidth]{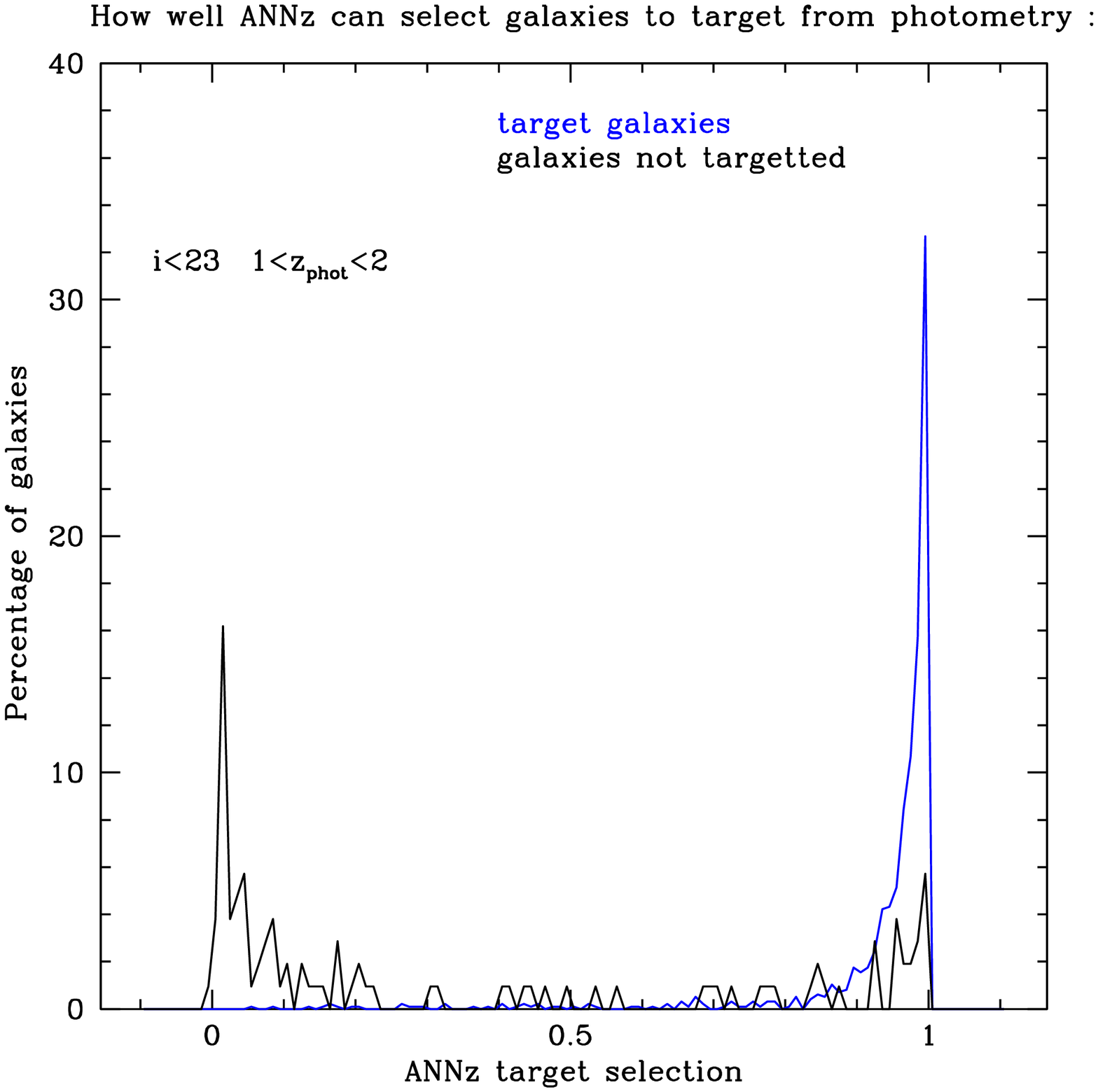}
  \caption{Histograms of ANNz target selection values for i$<$23 by deg2 
with different photometric redshift selection.  
The middle and right figures are the same as the left
figures with 0.5$<z_{phot}<$1 and 1$<z_{phot}<$2 for respectively the middle and right panels.}
\label{fig:nn_target}
\end{figure}
The blue line corresponds to the galaxies for which we will be able to measure a redshift,
the “target galaxies”; they have at least one emission line detected at 5 sigma 
and we assign them a value of 1. The black line corresponds to the galaxies that
do not have a strong emission line detectable, we assign them a value of 0.
Using 10000 galaxies as a training sample, we derive ANNz target selection values.

If photo-z selection is applied we can select galaxies at high
redshift with 95.5\% of success and galaxies at intermediate redshifts (between 0.5 and 1) with
88\% of success rate. DES photometry complemented with VHS photometry are assumed.

\begin{figure}
\resizebox{\hsize}{!}{\includegraphics{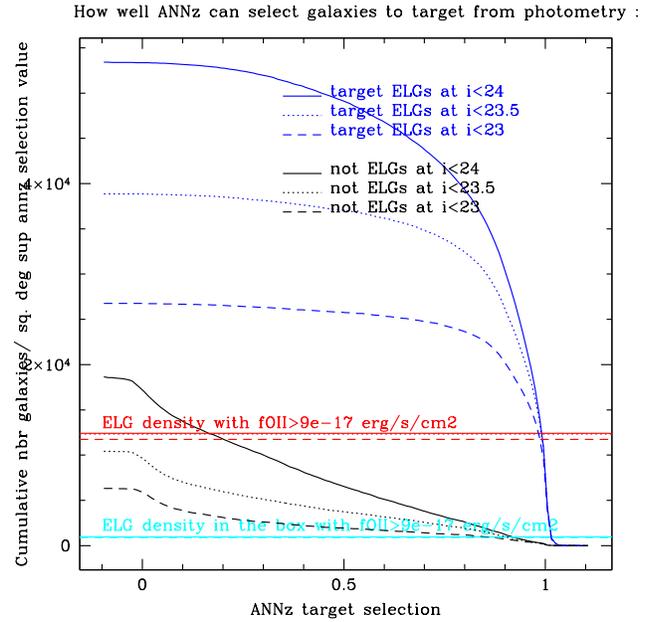}}
  \caption{Cumulative number of galaxies/deg2 as a function of the ANNz target selection
for ELGs.
Using a ANNz target criterion of 0.8, we are able to reach a SSR of 95\% at i$<$23 AB mag
targeting close to 75\% of all the galaxies at i$<$23. We will have a yield of 24000 galaxies/deg2.}
\label{fig:cumul_annz_elgs}
\end{figure}

To define a selection criterion from the ANNz values, we draw the cumulative number of
galaxies by deg2 as shown in \autoref{fig:cumul_annz_elgs}.
A criterion of 0.8 select most of emission line
galaxies and have a small contamination of weak or no emission line galaxies.

\subsection{Redshift distribution of color and NN selection for ELGs}
\begin{figure}
\resizebox{\hsize}{!}{\includegraphics{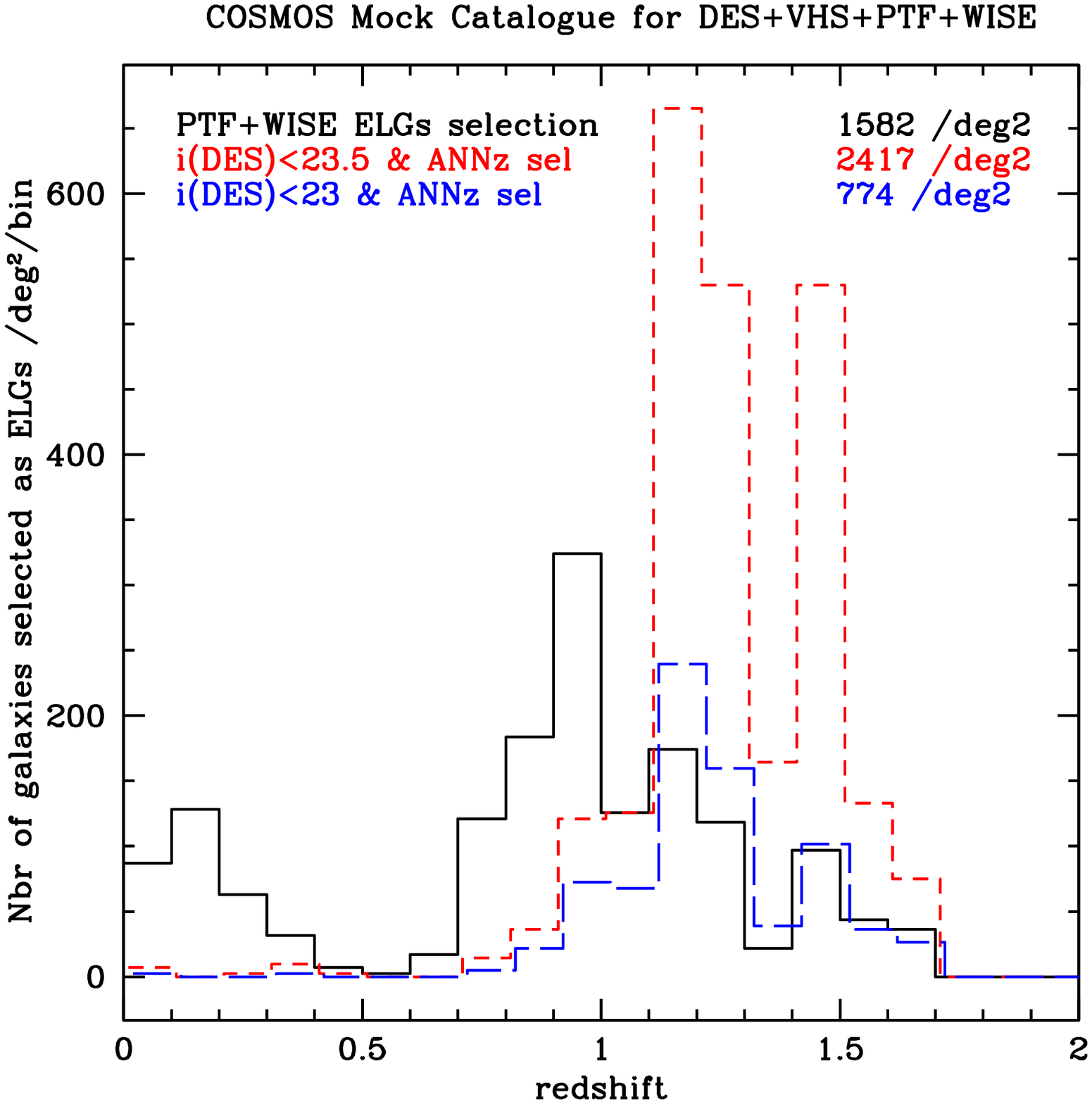}}
 \caption{Redshift distribution for the color-color selection defined in Figure
\autoref{fig:elg_col}. We add a selection function based on the photometric redshift
in order to select galaxies at redshift higher than 0.5.}
\label{fig:nz_elg}
\end{figure}

\autoref{fig:nz_elg} show ELGs redshifts distributions of the PTF+WISE
color selection, and the DES+VHS NN selection at i brighter than 23
and 23.5 in respectively black solid, long-dashed blue and small-dashed
red lines. The PTF+WISE curve include all the galaxies that are inside the
box regardless of their type. The DES+VHS NN selection shows
all the galaxies that have a NN value $>0.8$ and a selection
function based on the photoz. We shape the redshift distribution
in order to select all the high-redshift galaxies and have a smaller efficiency
for the galaxies at lower redshift. 
We aim at to have a redshift distribution as flat as possible.
Percival et al. shows that a flat redshift distribution gives the highest
dark energy constraints. 

In order to not be shot-noise limited, the target selection needs
to yield 80gal/deg$^2$ by 0.1 bin in redshift. With the photoz selection
function and the NN target selection, we will be able to study BAO
up to z$\approx$ 1.7 without being shot-noise limited. 

\subsection{Spectroscopic Success Rate of color and NN selection for ELGs}

\begin{figure}
\includegraphics[width=\columnwidth,height=0.9\columnwidth]{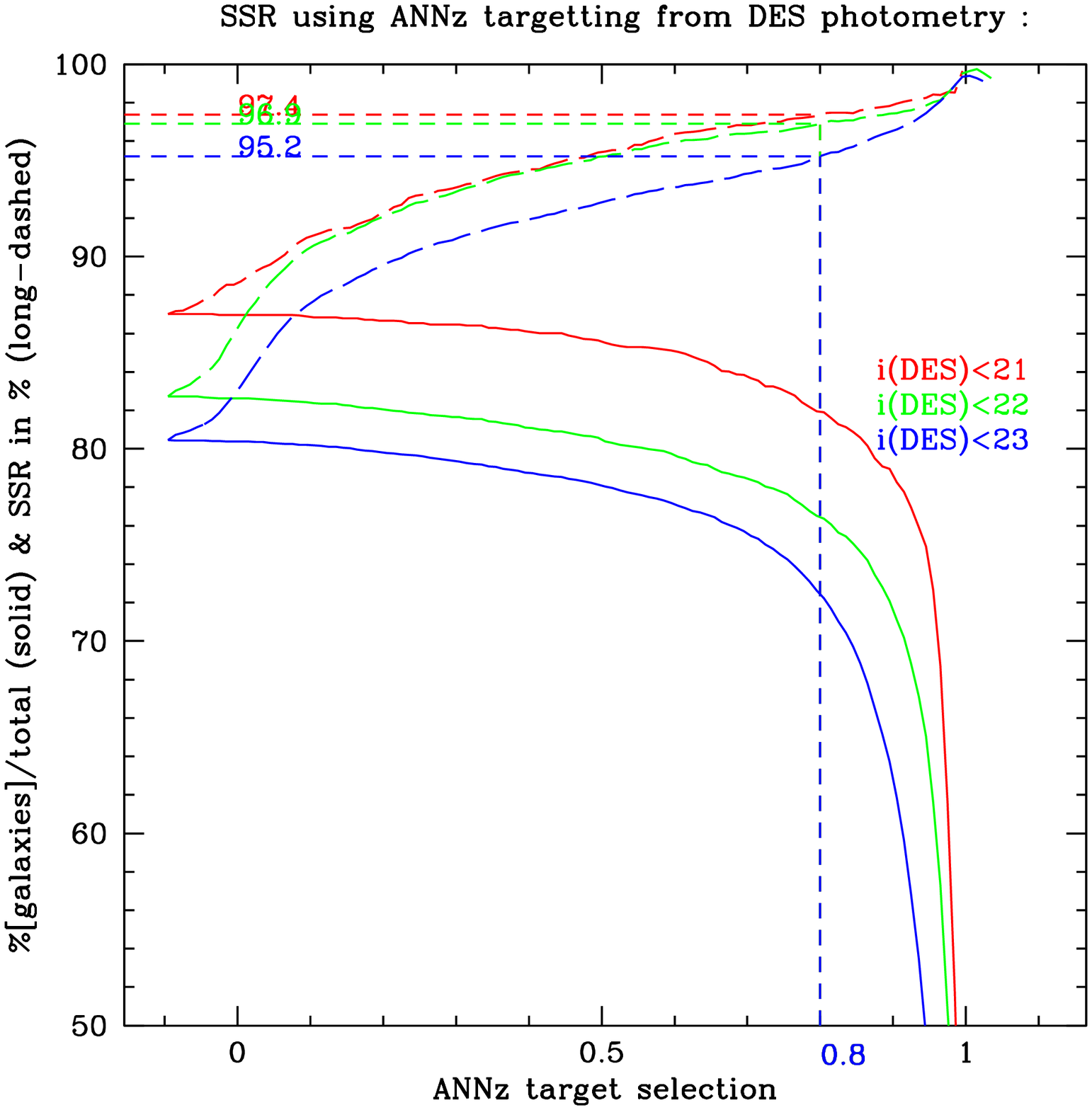}
\includegraphics[width=\columnwidth,height=0.9\columnwidth]{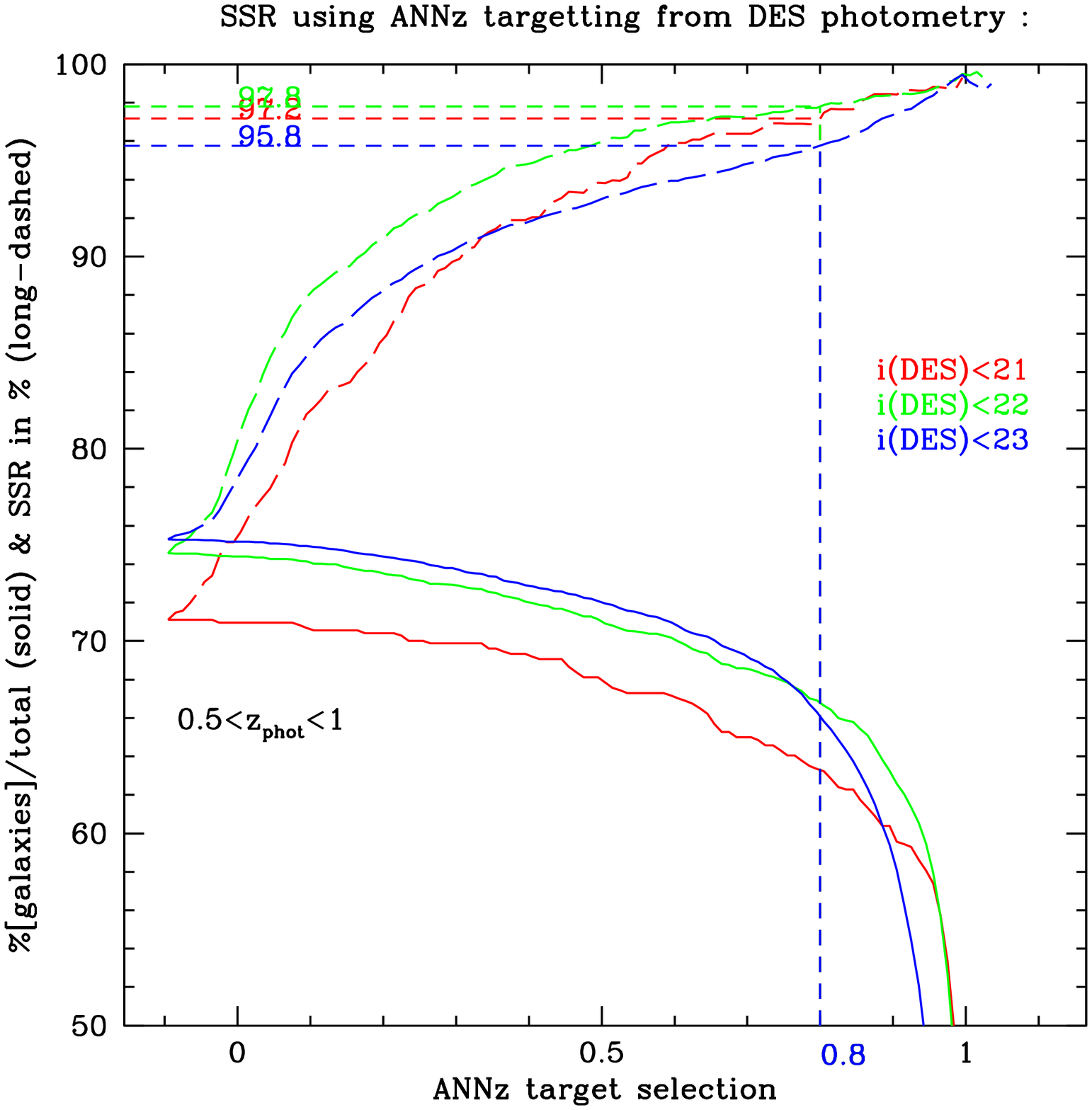}
\includegraphics[width=\columnwidth,height=0.9\columnwidth]{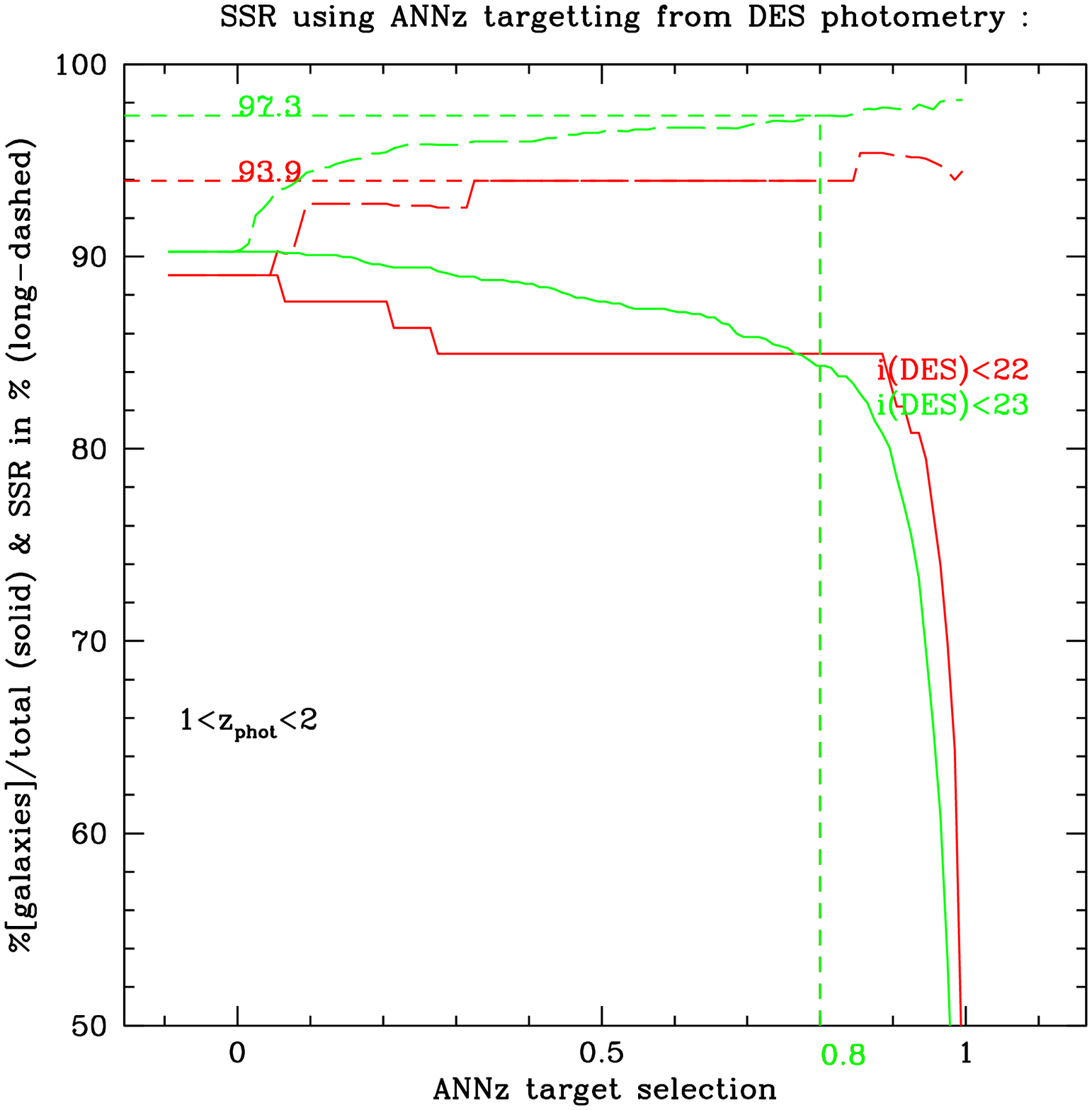}
  \caption{Spectroscopic Success Rate (dashed lines) and cumulative number of galaxies
(solid lines) as a function of ANNz target selection values for a magnitude i$<$23 in blue,
i$<$22 in green, and i$<$21 in red. The middle and right figures are the same as the left
figures with 0.5$<z_{phot}<$1 and 1$<z_{phot}<$2 for respectively the middle and right panels.}
\label{fig:ssr_elgs} 
\end{figure}

We then derive
the Spectroscopic Success Rate of ELGs for which we will be able to measure
a redshift as shown in \autoref{fig:ssr_elgs}. The dotted lines represent the SSR for different magnitude
cut while the solid lines represent the percentage of galaxies compared to the total
number of galaxies at the magnitude cut.

\section{Forecasts for Different Survey Strategy Assumptions}
\label{sec:strategy}

In this section and the next we present forecasts for our spectroscopic survey 
for a range of survey strategy and target selection options. Forecasts 
are made using the Fisher matrix formalism, described in section \ref{sec:FM}, 
then our choices for fiducial survey scenarios are outlined in section \ref{sec:survey_assumptions}. 
Results are presented in section \ref{sec:results}.

In this paper we concentrate on the effect of survey strategy on DE FoM \citep{DETF}, 
for a more detailed investigation of cosmological, astrophysical and systematic 
effects please see our companion paper \citet{kirkea_2013}.

\subsection{Fisher formalism background}
\label{sec:FM}

The forecasts in this section are calculated using the Fisher Matrix (FM) formalism \citep{Kendall77,Tegmark97}. The FM is calculated as
\begin{equation}
F_{\alpha\beta} = \sum^{l_{max}}_{l=l_{min}} \sum_{(i,j),(m,n)} \frac{\partial D_{ij}(l)}{\partial p_{\alpha} }{\rm Cov}^{-1} \left[ D_{ij}(l),D_{mn}(l) \right] \frac{\partial D_{mn}(l)}{\partial p_{\beta} }.
\end{equation}
where $D_{ij}$ is the data vector for the observables under consideration, 
$\frac{\partial D_{mn}(l)}{\partial p_{\beta}}$ are the derivatives of this data 
vector with respect to the cosmological parameters, ${\rm Cov} \left[ D_{ij}(l),D_{mn}(l) \right]$ 
is the covariance matrix of the data vector, calculated according to eqn. 39 of \citet{Joachimi09}. 

For these forecasts we use projected angular power spectra, $C_{ij}(l)$s, as our 
observables. Each survey we consider is broken into a number of tomographic redshift bins 
and the auto-/cross-correlations of each of these bins is considered. The subscript $ij$ labels 
a tomographic bin-pair.

We consider forecasts for a galaxy survey with spectroscopic quality redshifts. 
In this case the data vector is simply the galaxy-galaxy $C(l)$s, 
i.e. $D_{ij}(l) = C^{nn}_{ij}(l)$. We also consider forecasts for a 
DES-like WGL survey with photometric quality redshifts. In this case 
the data vector is shear-shear power spectrum, i.e. $D_{ij}(l) = C^{\epsilon\epsilon}_{ij}(l)$. 
We consider combinations of these two surveys when there is no common overlap and 
when the two surveys fully overlap. In the no-overlap case the observables are 
independent and the separate FMs can simply be summed together. 
In the case of overlapping surveys we need to consider cross-correlations 
between the galaxy and shear observables. We calculate a single FM with data 
vector $D_{ij}(l) = \{ C^{\epsilon\epsilon}_{ij}(l) C^{n\epsilon}_{ij}(l) C^{nn}_{ij}(l) \}$, 
where $C^{n\epsilon}_{ij}(l)$ are the galaxy-shear cross-correlations.

A general $C(l)$ is the projection of two window functions
\begin{equation}
C_{\alpha\beta}^{i,j}(l) = 4 \pi \int\Delta^{2}(k)W_{\alpha}^{i}(k)W_{\beta}^{j}(k)\frac{dk}{k}.
\end{equation}
$W^{i}_{\alpha}(k)$ is the window function for tomographic bin $i$ and observable $\alpha$. In this document we consider three different $C(l)$ observables: $C_{nn}^{ij}(l)$, the galaxy-galaxy correlation (elsewhere called galaxy clustering), $C_{\epsilon\epsilon}^{ij}(l)$, the shear-shear correlation from WGL and $C_{n\epsilon}^{ij}(l)$, the galaxy-shear cross-correlation. 

The weight function for galaxies is
\begin{equation}
\begin{split}
W_{n}^{i}(l,k) &= \beta\int n(z(\chi)) \Bigl[ \frac{(2l^{2}+2l-1)}{(2l+3)(2l-1)}j_{l}(k\chi) - \\
&\frac{l(l-1)}{(2l-1)(2l+1)}j_{l-2}(k\chi) - \frac{(l+1)(l+2)}{(2l+1)(2l+3)}j_{l+2}(k\chi) \Bigr] d\chi,
\end{split}
\end{equation}
where $n^{i}(z)$ is the galaxy redshift distribution of tomographic bin $i$, $j_{l}(kz)$ 
is a bessel function and this expression includes the RSD formalism of \citet{Padmanabhan05,Fisher94}. 

For cosmic shear the weight function is 
\begin{equation}
W_{\epsilon}^{i}(l,k) = \int \frac{3H_{0}^{2}\Omega_m}{2c^2}\frac{\chi}{a(\chi)}\int_{\chi_{hor}}^{\chi} p^{(i)}(\chi')\frac{\chi'-\chi}{\chi'} d\chi' j_{l}(k,\chi) d\chi.
\end{equation}

We constrain the set of cosmological parameters $\left\{ \Omega_m, w_0, w_a, h, \sigma_8, \Omega_b, n_s \right\}$ which take the values $\Omega_m=0.25$, $w_0=-1$, $w_a=0$, $h=0.7$, $\sigma_8=0.8$, $\Omega_b=0.05$, $n_s=1$. All results assume a flat $\Lambda$CDM cosmology. \\

\subsubsection{Survey Scenario Assumptions}
\label{sec:survey_assumptions}

For the purpose of these forecasts we describe two types of survey: 
a photometric WGL survey with photo-z quality redshifts and a spectroscopic 
survey with spec-z quality redshifts. For the spec-z survey we vary a number 
of properties to assess the impact of survey strategy and target selection. 
These choices are detailed below. We consider the properties of the photo-z 
survey to be fixed at the DES-reference values because we are primarily 
interested in the effect of choices in spec-z survey strategy on the combined photo-z + spec-z constraints.

The photometric survey is assumed to cover 5,000deg$^2$ with a galaxy number 
density of 10 arcmin$^{-2}$ ($\sim$ 300 million galaxies in total) and a redshift distribution given by the Smail-type
$n(z)$,
\begin{equation}
n(z)=z^{\alpha}\exp{\left(- \left( \frac{z}{z_{0}} \right)^{\beta}\right)}
\end{equation}
with $\alpha=2$, $\beta=1.5$, $z_0=0.9/\sqrt{2}$, divided into 5 tomographic bins 
with equal number density out to redshift 3 and normalised so 
that $\int n(z) dz = 1$. A Gaussian photometric redshift 
error of $\sigma_z = 0.07(1+z)$ was applied with no catastrophic outliers in redshift. 

The photo-z survey has an area of 5,000deg$^2$ targetting $\sim$100 million 
galaxies (though both of these are varied, see below). The galaxy redshift distribution 
depends on choices of survey strategy and target selection described below but is 
assumed to cover redshifts from 0.4 to 1.7 and is split into 20 tomographic bins of 
equal z-range. We marginalise over our standard 7 cosmological parameters plus a galaxy bias 
model, $b_{g}(k,z) = A_{b_{g}}Q(k,z)$, which allows a variable overall amplitude term and 
a scale/redshift dependent grid parameterised with $2 \times 2$ nodes in k/z. The fiducial values of the amplitude and scale/redshift nodes is unity. Our fiducial bias model is therefore also unity but our variable amplitude and grid nodes allow us to explore a range of flexible z- and k-dependent bias forms. These z/k-dependent biases are not tied to any specific models based on physical arguments or simulations, we are aiming in this case for maximum generality as a stringent test of our constraining power.

In section \ref{sec:results} we present forecasts for our spectroscopic survey alone 
and in combination with our photometric survey. Each spectroscopic forecast 
assumes a particular exposure time (from 20min to 50min) which, in turn, determines 
the total number of targets available in the telescope field of view, assumed to be 3deg$^2$. 
A longer exposure time allows fainter targets to be resolved hence a longer target list. 
Possible LRG and ELG targets are calculated independently. The spectrograph is assumed 
to have 3000 fibres covering the field of view allowing a maximum number of 1333 targets per 
square degree to be captured per exposure. If the number of available targets is 
less than this we assume the target list is saturated. If the number of possible 
targets is greater than the number of available fibres then a sub-set of 1333 
galaxies are assumed to be captured at some specified ratio of LRG/ELG targets. 
Finally, for each forecast we assumed a total survey area (5000deg$^2$ - 15000deg$^2$) 
and the number of times the area is tiled i.e. is each square degree of survey subject to 
a single exposure or does the telescope revisit multiple times. These multiple visits are 
useful when there are more available targets than fibres, allowing the excess to be 
captured on return visits. The combination of survey area, exposure time and number 
of tilings determines the total survey time in nights, assuming 8 hours observing per night. 
We assume a 90\% success rate for spectral acquisition. 

To summarise, the variables we consider in forecasting our spectroscopic survey are: 
i) exposure time, ii) LRG/ELG ratio, iii) number of tilings and iv) survey area.

\section{Results}
\label{sec:results}

Here we present the results of our forecasts for a spec-z survey as a function of 
survey strategy and target selection choices. Throughout this section we will quote values for the Dark Energy Figure of Merit (DE FoM), based on the Dark Energy Task Force \citep{DETF} definition
\begin{equation}
\textrm{FoM}_{\textrm{DE}} = \frac{1}{4\sqrt{\textrm{det}(F^{-1})_{\textrm{DE}}}},
\end{equation}
where the subscript $\textrm{DE}$ denotes the $2 \times 2$ sub-matrix of the inverse FM that corresponds to the entries for the equation of state of DE parameters, $w_0$ and $w_a$. Note that different prefactors to this equation exist in the literature. We use the factor of $1/4$ for consistency with related papers \citep{Bridle07,Joachimi09}.

This sort of FoM-based optimisation of surveys has been carried out numerous times including the examples of \citet{Bassett05,Parkinson07,Yamamoto07,Parkinson10,Paykari13,kirkea_2011,Amara07}.

\subsection{FoM as a function of LRG/ELG ratio}

\begin{figure}
\resizebox{\hsize}{!}{\includegraphics{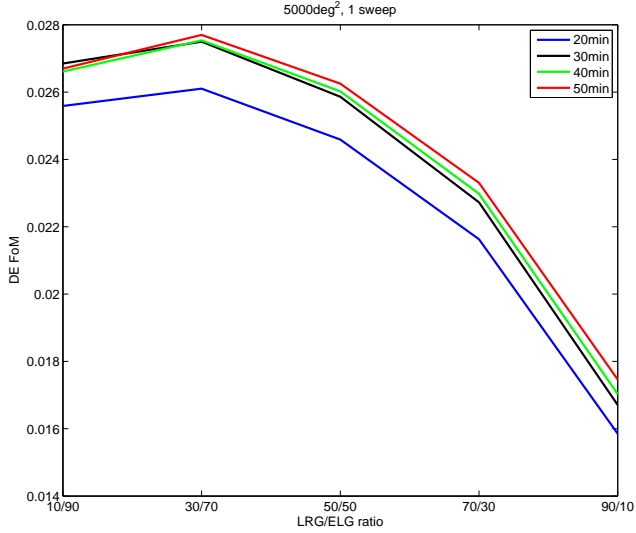}}
\caption{DE FoM as a function of LRG/ELG ratio for a range of exposure times. 5000deg$^2$ survey is assumed at all times and the area is tiled once. 7 cosmological parameters are marginalised over plus 5 galaxy bias parameters. Standard k-cuts assumed. *Possibly add survey time to legend*}
\label{fig:FoM_elg_lrg_ratio}
\end{figure}

\autoref{fig:FoM_elg_lrg_ratio} shows the effect of selecting different ratios of 
LRG/ELG targets for a range of exposure times. Survey area is assumed fixed at 5000deg$^2$ 
and each forecast assumes one tiling of that area i.e. number density of galaxies is nowhere greater than 1333deg$^{-2}$. 

It is clear that the trend with LRG/ELG ratio is robust across exposure times. 
The shortest exposure time, 20min, performs significantly worse than the others because 
such a short exposure with one tiling fails to saturate the number of available fibres with targets, 
resulting in significantly lower number densities. Even so the trend with target type is the same. 

There is a clear peak at 30\% LRG/ 70\% LRG, while a strategy that focuses primarily on 
LRGs is clearly sub-optimal. ELGs predominate at high redshift so sacrificing them 
limits the volume of the survey. The inclusion of 30\% LRGs seems sufficient to maintain 
good coverage across the z-range. Note we have not assumed each population to have 
a different galaxy bias. In practice ELGs are thought to be more strongly biased than LRGs. 
If biasing is equally well understood then more strongly biased populations are a more 
powerful probe of cosmology due to their larger signal to noise. This effect would enhance 
the trend seen on our results. 

\subsection{Area versus depth for fixed survey time}

\begin{figure}
\resizebox{\hsize}{!}{\includegraphics{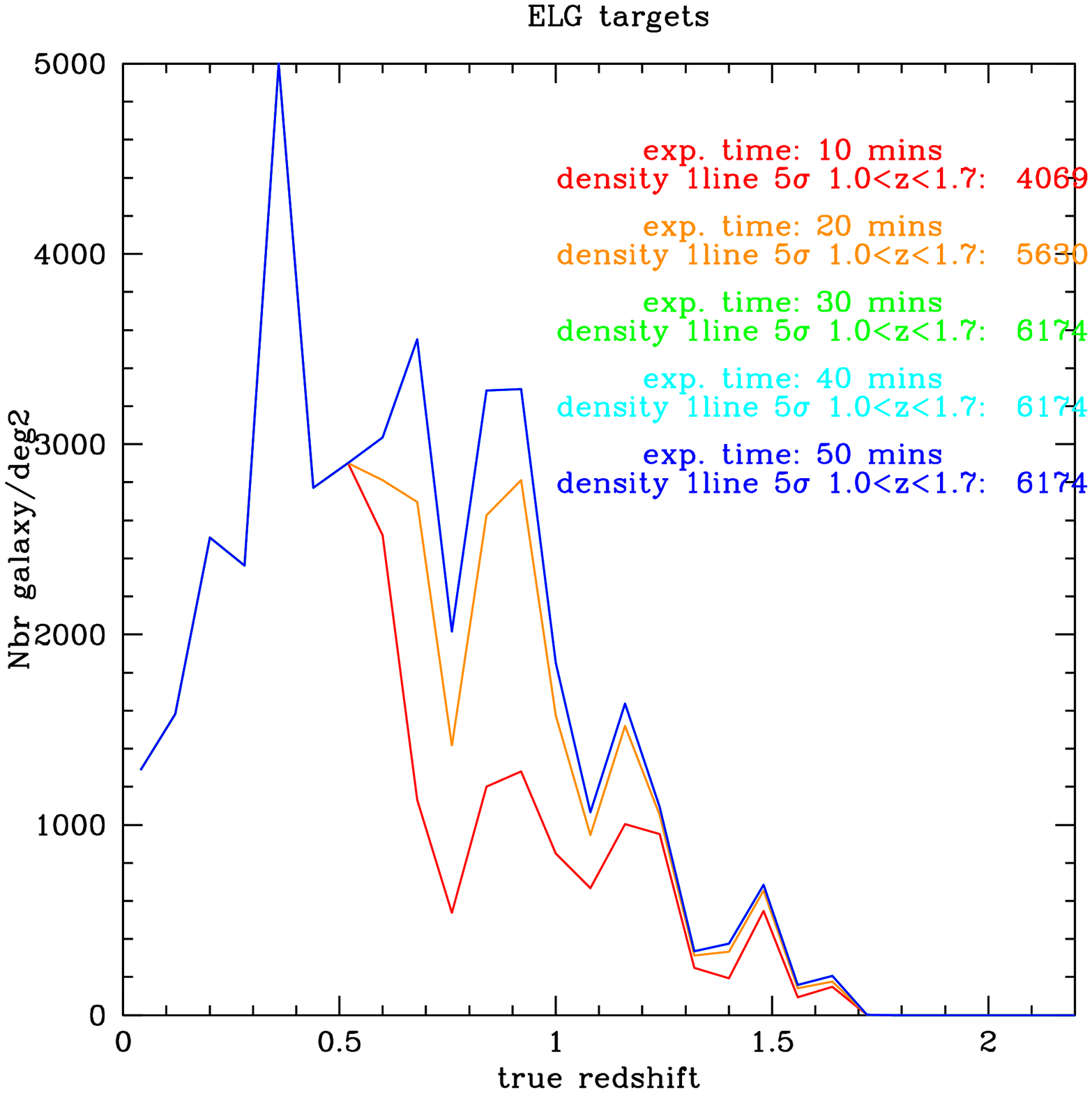}}
\resizebox{\hsize}{!}{\includegraphics{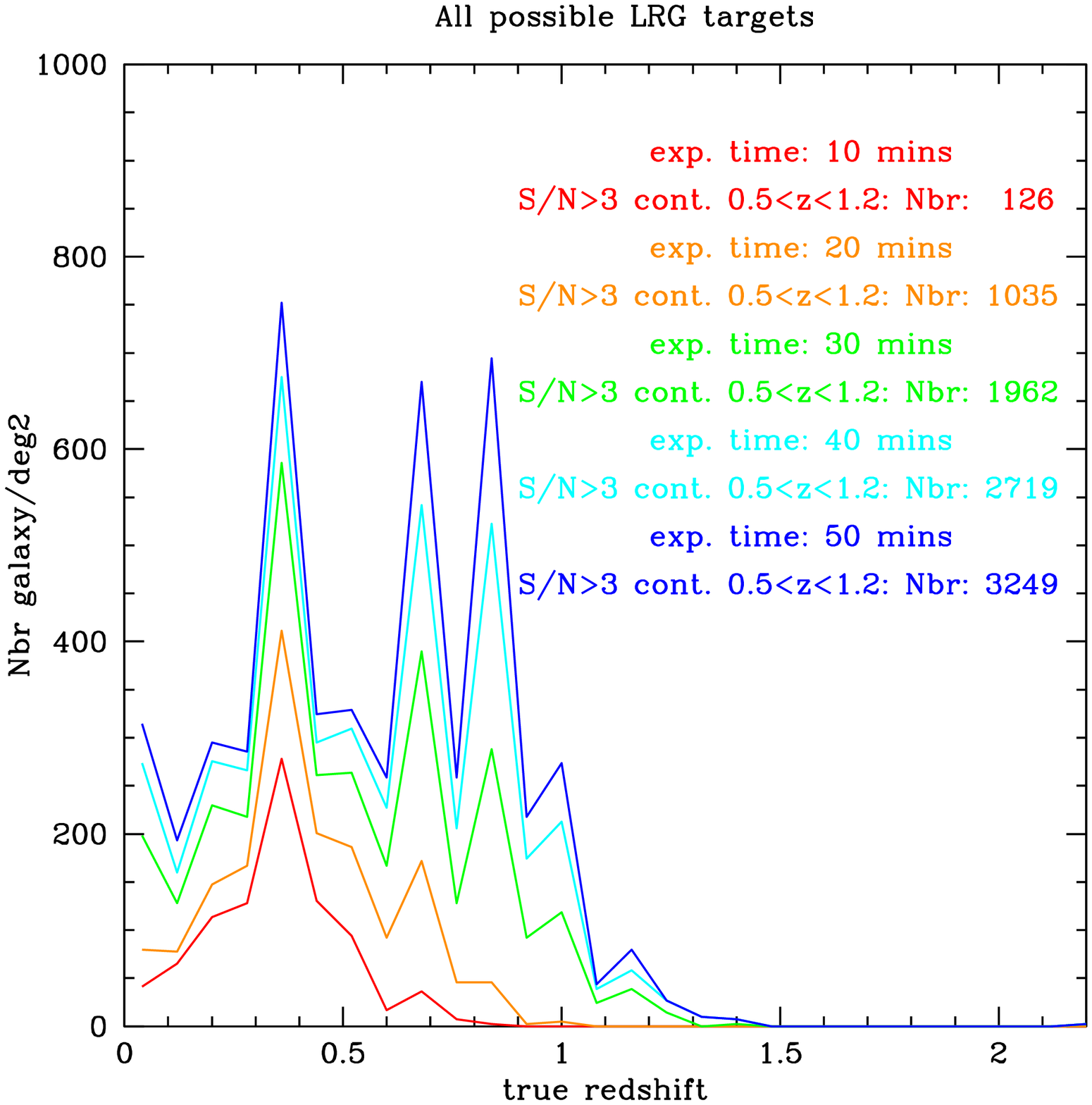}}
\caption{(Top panel) Redshift histograms for ELGs brighter than i $\sim$ 23.5. 
The different lines corresponds to successful redshifts obtained for different observation times. 
We note that the curves saturates after 30mins exposure as our photometric target selection 
is such that most emission lines are obtained.(Bottom panel) Redshifts histograms for LRGS brighter than
i$\sim$ 22.}
\label{fig:nz_elg_wvl}
\end{figure}

\begin{figure}
\resizebox{\hsize}{!}{\includegraphics{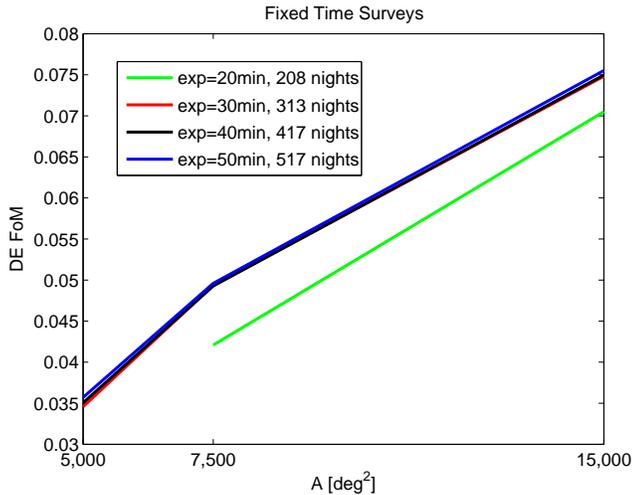}}
\caption{DE FoM as a function of exposure time for a 5000deg$^2$ survey where the area is tiled once. Where there are more targets than fibres an ELG/LRG ratio of 30/70 is assumed, otherwise the available target list is saturated. Total survey time in nights is noted beside each forecast point.  Seven cosmological parameters are marginalised over plus 5 galaxy bias parameters. Standard k-cuts assumed.}
\label{fig:FoM_fixedtime}
\end{figure}

\autoref{fig:FoM_fixedtime} shows DE FoM as a function of survey area for a range of 
exposure times. The total survey time in nights is noted in the legend. To keep total survey 
time constant for each exposure time we trade off area and number of tilings. We assume 
3 tilings for each 5000deg$^2$ forecast, 2 tilings for the 7,500deg$^2$ forecasts and 
a single tiling for the 15,000deg$^2$ forecasts. Some forecasts will be limited by the 
number of available fibres, 1,333deg$^{-2}$ per tiling, where this is less than the 
number of available targets. We do not forecast a 20min exposure time survey over 5000deg$^2$ 
with 3 tilings because the available target list is already saturated with 2 tilings so 
an additional pass would add nothing to a survey design.

More area is clearly beneficial, even at the cost of number density. This agrees 
with the findings of \citet{Amara07} and \citet{MGPaper2}. 20min exposure times perform 
significantly worse due to their low number density. Above this level exposure time has 
little effect on the results because number density is limited by available fibres, 
changes in galaxy redshift distribution have a small effect, sub-dominant compared to the impact of number density/area. 

\subsection{Fixed Area, single survey sweep}

In this section we forecast DE FoM for a 5000deg$^2$ survey as a function of exposure time, 
assuming the whole area is tiled once. Note that this means that, for exposure times of 
20min and above, our acquired targets are limited by the 1,333 fibres per deg$^2$ available 
in our presumed instrument design.

\begin{figure}
\resizebox{\hsize}{!}{\includegraphics{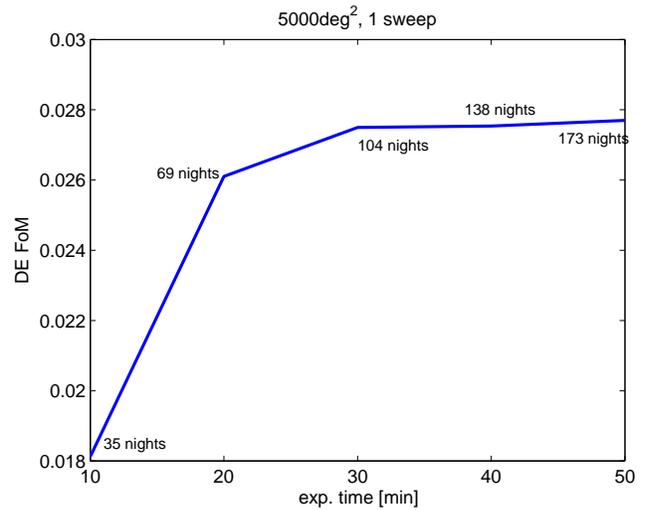}}
\caption{DE FoM as a function of exposure time. We assume a 5000deg$^2$ survey area which is tiled once. Total survey time in nights is noted next to each forecast. Where there are more targets than fibres an ELG/LRG ratio of 30/70 is assumed, otherwise the available target list is saturated. Seven cosmological parameters are marginalised over plus 5 galaxy bias parameters. Standard k-cuts assumed.}
\label{fig:FoM_fixedA_1sweep}
\end{figure}

While DE FoM improves with increased exposure time, the majority of the available information 
is obtained by 20min exposure, with almost everything captured by 30min. This is due to the 
fact that number density, limited by number of fibres, is constant for 20min and above, 
any remaining benefit comes from the more even redshift distribution afforded by longer exposure times.

\subsection{Fixed Area, full target list}

\begin{figure}
\resizebox{\hsize}{!}{\includegraphics{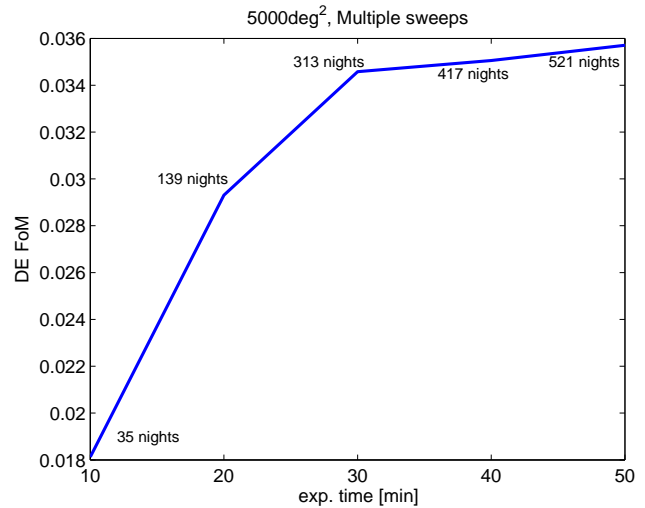}}
\caption{DE FoM as a function of exposure time. We assume a 5000deg$^2$ survey area which is tiled as many times as necessary to saturate the target list. This corresponds to 1 tiling for 20min exposures, 2 tilings for 20min and 30min exposures and 3 tilings for 40min and 50min exposures. Total survey time in nights is noted next to each forecast. Seven cosmological parameters are marginalised over plus 5 galaxy bias parameters. Standard k-cuts assumed.}
\label{fig:FoM_fixedA_manysweeps}
\end{figure}

When we fix our survey area to 5000deg$^2$ and allow enough tilings to saturate the 
target list for each exposure time we forecast the DE FoMs shown in 
\autoref{fig:FoM_fixedA_manysweeps}. Increased exposure time here leads to drastically increased 
total survey time as, not only does each individual pointing take more time, but entire 
5000deg$^2$ tilings must be added if the number of available targets is greater than the 
number of available fibres. in practice we need 1 tiling for 10min exposure time, 2 
tilings for 20min \& 30min and 3 tilings for 40min \& 50min.

The increased number density provided by exposure times of over 30min adds negligibly to 
the constraining power of the survey. A survey of $\sim 300$ nights captures almost all the 
available information. Note that there is some redundancy here as an entire extra tiling 
may not be the most profitable use of telescope time to catch a small number of targets 
missed due to lack of fibres on the previous tiling.

\subsection{FoM as a function of wavelength}

In this subsection, we investigate how the FoM changes depending on the maximum wavelength of the
spectrograph. We can clearly see from the target selection plots that at high redshift most if not all
the targets are obtained via the OII lines. Therefore there is a one-to-one relation 
between the maximum redshift that is achieved by our expereiemnt and the OII line dropping out of the
range available by the spectrograph. We made simulations for the redshift histograms that can 
are obtainned spectroscopically for different detector cut offs and plotted the resulting distributions in 
\autoref{fig:nz_elg_wvl}.

\begin{figure}
\resizebox{\hsize}{!}{\includegraphics{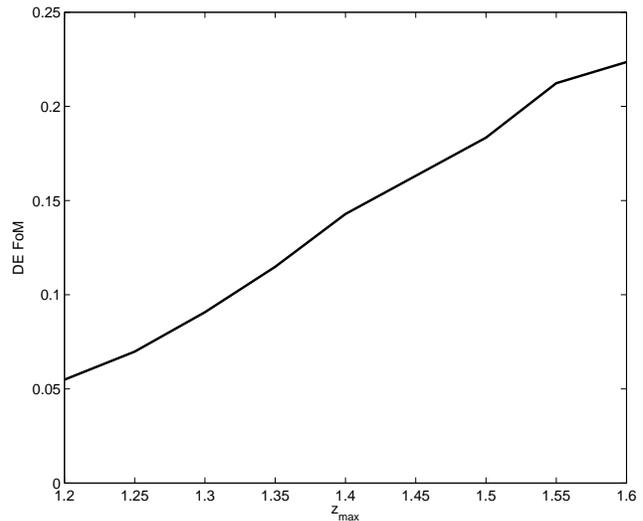}}
\caption{FoM as a function of the maximum redshift achievable by the spectroscopic instrument via the OII line.}
\label{fig:fom_elg_wvl}
\end{figure}

There is a significant amount of cosmic volume available at these redshifts and hence this redshift cutoff
can significantly reduce the FOMs for Dark Energy recovery. We have calculated these FOM removing all information
that is present at such redshifts and plotted the results in \autoref{fig:fom_elg_wvl}. For reference the OII
line is present at 3727 $\AA$ so a spectrograph which cuts off at 8000 $\AA$ would be able to detect OII out to $z \sim 1.14$
and a cutoff of 1 micron would correspond to OII at redshift 1.68. We can see from \autoref{fig:fom_elg_wvl}
that the FOMs can cange by a factor of 
four by including these extra galaxies at high redshifts.

\begin{figure}
\resizebox{\hsize}{!}{\includegraphics{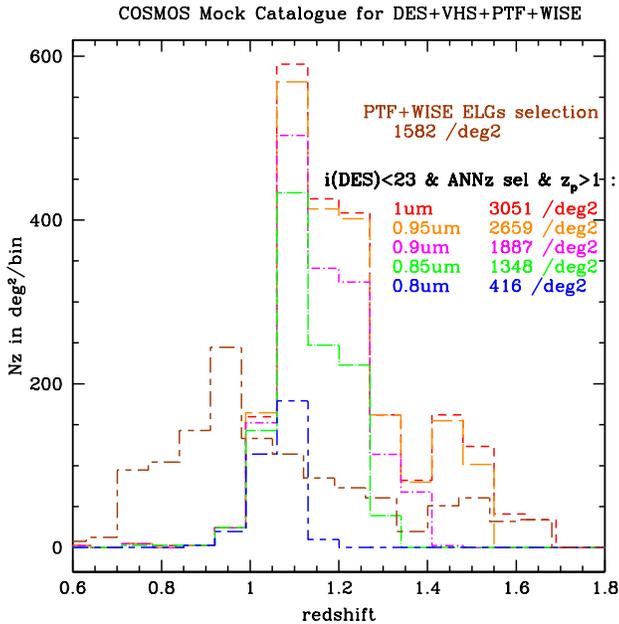}}
\caption{Nz of ELGs as a function of the spectrograph wavelength.}
\label{fig:nz_elg_wvl}
\end{figure}

\section{Conclusion}
In this paper we have produced a full forecast pipeline from realistic mock catalogues up to the DE FoM. 
With this pipeline we investigate several scenarios of spectroscopic survey design and seek to optimise that design in several ways.  
Using a spectroscopic instrument simulation similar to DESpec, BigBOSS or DESI, we first simulated
a target selection strategy using shallow and deep photometry to cover
the available range of current surveys such as DES, LSST, panSTARRS, PTF, WISE, VISTA.
We have compared conservative color-color target selections and more complex ones based
on neural network optimisation. 
We compare redshift distributions, success rate and completness of the target selections and
find a very good results for the neural network selection. 

We choose to target a mix of emission line galaxies and lyman red galaxies to cover the redshift
range from 0.5 up to 1.7. For LRGs brighter than $i\sim22$, we reach an SSR of 98\% 
success for the neural network against 48 and 53\% for
color-color selections using respectively PTF-WISE and DES-VHS photometry. 
For the ELGs, we reach an SSR higher than 95\% at magnitude i brighter than 23 for redshifts
up to 1.3.

We thus show that having deep multicolor photometry for the target selection allows us to increase the number 
density of LRGs and ELGs targets by a factor of 2 to 3. It will also improve the efficiency
and success rate of the target selection. 
Using the redshift distribution and target densities we find with the neural network selection, 
we give a first estimation of spectroscopic survey strategy. 
We look at tradeoff between ELG/LRG ratio survey area versus depth using FoM. The assumptions which
go into the FoM calculation will be described in more detail in \citet{kirkea_2013}.  
We find a ratio of 30\% LRG 70\% ELG to be optimal when producing figures of merit for dark energy.
Concerning the survey strategy, an exposure time shorter than 20mins is clearly sub-optimal and 
there is no significant benefit beyond 30mins. The optimal choice between these exposure times
depends on survey strategy choices, area versus completness of the target list. 
The procedure presented in this paper is a general approach. We are able to use it to optimise the target selectio
and survey strategy for any future spectroscopic and overlapping photometric surveys

\bibliographystyle{mn2e}
\bibliography{biblio_mnras}

\begin{thebibliography}{}

\bibitem[\protect\citeauthoryear{{Abdalla}, {Annis}, {Bacon}, {Bridle},
  {Castander}, {Colless}, {DePoy}, {Diehl}, {Eriksen}, {Flaugher}, {Frieman} \&
  {Gaztanaga}}{{Abdalla} et~al.}{2012}]{Abdalla12}
{Abdalla} F.,  {Annis} J.,  {Bacon} D.,  {Bridle} S.,  {Castander} F.,
  {Colless} M.,  {DePoy} D.,  {Diehl} H.~T.,  {Eriksen} M.,  {Flaugher} B.,
  {Frieman} J.,    {Gaztanaga} E. e.~a.,  2012, ArXiv e-prints

\bibitem[\protect\citeauthoryear{{Abdalla}, {Amara}, {Capak}, {Cypriano},
  {Lahav} \& {Rhodes}}{{Abdalla} et~al.}{2008}]{Abdalla08}
{Abdalla} F.~B.,  {Amara} A.,  {Capak} P.,  {Cypriano} E.~S.,  {Lahav} O.,
  {Rhodes} J.,  2008, \mnras, 387, 969

\bibitem[\protect\citeauthoryear{{Abdalla}, {Mateus}, {Santos}, {Sodr{\`e}}
  Jr., {Ferreras} \& {Lahav}}{{Abdalla} et~al.}{2008}]{Abdalla08nn}
{Abdalla} F.~B.,  {Mateus} A.,  {Santos} W.~A.,  {Sodr{\`e}} Jr. L.,
  {Ferreras} I.,    {Lahav} O.,  2008, \mnras, 387, 945

\bibitem[\protect\citeauthoryear{{Albrecht}, {Bernstein}, {Cahn}, {Freedman},
  {Hewitt}, {Hu}, {Huth}, {Kamionkowski}, {Kolb}, {Knox}, {Mather}, {Staggs} \&
  {Suntzeff}}{{Albrecht} et~al.}{2006}]{DETF}
{Albrecht} A.,  {Bernstein} G.,  {Cahn} R.,  {Freedman} W.~L.,  {Hewitt} J.,
  {Hu} W.,  {Huth} J.,  {Kamionkowski} M.,  {Kolb} E.~W.,  {Knox} L.,  {Mather}
  J.~C.,  {Staggs} S.,    {Suntzeff} N.~B.,  2006, ArXiv Astrophysics e-prints

\bibitem[\protect\citeauthoryear{{Amara} \& {R{\'e}fr{\'e}gier}}{{Amara} \&
  {R{\'e}fr{\'e}gier}}{2007}]{Amara07}
{Amara} A.,  {R{\'e}fr{\'e}gier} A.,  2007, \mnras, 381, 1018

\bibitem[\protect\citeauthoryear{{Banerji}, {Abdalla}, {Lahav} \&
  {Lin}}{{Banerji} et~al.}{2008}]{Banerji08}
{Banerji} M.,  {Abdalla} F.~B.,  {Lahav} O.,    {Lin} H.,  2008, \mnras, 386,
  1219

\bibitem[\protect\citeauthoryear{{Bassett}, {Parkinson} \& {Nichol}}{{Bassett}
  et~al.}{2005}]{Bassett05}
{Bassett} B.~A.,  {Parkinson} D.,    {Nichol} R.~C.,  2005, \apjl, 626, L1

\bibitem[\protect\citeauthoryear{{Blake}, {Brough}, {Colless}, {Contreras},
  {Couch}, {Croom}, {Croton}, {Davis}, {Drinkwater}, {Forster} \&
  {Gilbank}}{{Blake} et~al.}{2012}]{Blake12}
{Blake} C.,  {Brough} S.,  {Colless} M.,  {Contreras} C.,  {Couch} W.,  {Croom}
  S.,  {Croton} D.,  {Davis} T.~M.,  {Drinkwater} M.~J.,  {Forster} K.,
  {Gilbank} D. e.~a.,  2012, \mnras, 425, 405

\bibitem[\protect\citeauthoryear{{Blanton} \& {Roweis}}{{Blanton} \&
  {Roweis}}{2007}]{Blanton07}
{Blanton} M.~R.,  {Roweis} S.,  2007, \aj, 133, 734

\bibitem[\protect\citeauthoryear{{Bridle} \& {King}}{{Bridle} \&
  {King}}{2007}]{Bridle07}
{Bridle} S.,  {King} L.,  2007, New Journal of Physics, 9, 444

\bibitem[\protect\citeauthoryear{{Bruzual} \& {Charlot}}{{Bruzual} \&
  {Charlot}}{2003}]{Bruzual03}
{Bruzual} G.,  {Charlot} S.,  2003, \mnras, 344, 1000

\bibitem[\protect\citeauthoryear{{Capak}, {Aussel}, {Ajiki}, {McCracken},
  {Mobasher}, {Scoville}, {Shopbell} \& {Taniguchi}}{{Capak}
  et~al.}{2007}]{Capak07}
{Capak} P.,  {Aussel} H.,  {Ajiki} M.,  {McCracken} H.~J.,  {Mobasher} B.,
  {Scoville} N.,  {Shopbell} P.,    {Taniguchi} Y. e.~a.,  2007, \apjs, 172, 99

\bibitem[\protect\citeauthoryear{{Capak}, {Aussel}, {Ajiki}, {McCracken},
  {Mobasher}, {Scoville}, {Shopbell} \& {Taniguchi}}{{Capak}
  et~al.}{2008}]{Capak08}
{Capak} P.,  {Aussel} H.,  {Ajiki} M.,  {McCracken} H.~J.,  {Mobasher} B.,
  {Scoville} N.,  {Shopbell} P.,    {Taniguchi} Y. e.~a.,  2008, VizieR Online
  Data Catalog, 2284, 0

\bibitem[\protect\citeauthoryear{{Coe}, {Ben{\'{\i}}tez}, {S{\'a}nchez}, {Jee},
  {Bouwens} \& {Ford}}{{Coe} et~al.}{2006}]{Coe06}
{Coe} D.,  {Ben{\'{\i}}tez} N.,  {S{\'a}nchez} S.~F.,  {Jee} M.,  {Bouwens} R.,
     {Ford} H.,  2006, \aj, 132, 926

\bibitem[\protect\citeauthoryear{{Collister} \& {Lahav}}{{Collister} \&
  {Lahav}}{2004}]{Collister04}
{Collister} A.~A.,  {Lahav} O.,  2004, \pasp, 116, 345

\bibitem[\protect\citeauthoryear{{Dawson}, {Schlegel}, {Ahn}, {Anderson},
  {Aubourg}, {Bailey} \& {Barkhouser}}{{Dawson} et~al.}{2013}]{Dawson13}
{Dawson} K.~S.,  {Schlegel} D.~J.,  {Ahn} C.~P.,  {Anderson} S.~F.,  {Aubourg}
  {\'E}.,  {Bailey} S.,    {Barkhouser} R.~H. e.~a.,  2013, \aj, 145, 10

\bibitem[\protect\citeauthoryear{{de Jong}, {Bellido-Tirado}, {Chiappini},
  {Depagne}, {Haynes}, {Johl} \& {Schnurr}}{{de Jong} et~al.}{2012}]{4MOST}
{de Jong} R.~S.,  {Bellido-Tirado} O.,  {Chiappini} C.,  {Depagne} {\'E}.,
  {Haynes} R.,  {Johl} D.,    {Schnurr} O. e.~a.,  2012, in Society of
  Photo-Optical Instrumentation Engineers (SPIE) Conference Series Vol.~8446 of
  Society of Photo-Optical Instrumentation Engineers (SPIE) Conference Series,
  {4MOST: 4-metre multi-object spectroscopic telescope}

\bibitem[\protect\citeauthoryear{{Drinkwater}, {Jurek}, {Blake}, {Woods},
  {Pimbblet}, {Glazebrook}, {Sharp}, {Pracy}, {Brough}, {Colless}, {Couch} \&
  {Croom}}{{Drinkwater} et~al.}{2010}]{Drinkwater10}
{Drinkwater} M.~J.,  {Jurek} R.~J.,  {Blake} C.,  {Woods} D.,  {Pimbblet}
  K.~A.,  {Glazebrook} K.,  {Sharp} R.,  {Pracy} M.~B.,  {Brough} S.,
  {Colless} M.,  {Couch} W.~J.,    {Croom} S.~M. e.~a.,  2010, \mnras, 401,
  1429

\bibitem[\protect\citeauthoryear{{Eisenstein}, {Annis}, {Gunn}, {Szalay},
  {Connolly}, {Nichol} \& {Bahcall}}{{Eisenstein} et~al.}{2001}]{Eisenstein01}
{Eisenstein} D.~J.,  {Annis} J.,  {Gunn} J.~E.,  {Szalay} A.~S.,  {Connolly}
  A.~J.,  {Nichol} R.~C.,    {Bahcall} N.~A. e.~a.,  2001, \aj, 122, 2267

\bibitem[\protect\citeauthoryear{{Fisher}, {Scharf} \& {Lahav}}{{Fisher}
  et~al.}{1994}]{Fisher94}
{Fisher} K.~B.,  {Scharf} C.~A.,    {Lahav} O.,  1994, \mnras, 266, 219

\bibitem[\protect\citeauthoryear{{Gazta{\~n}aga}, {Eriksen}, {Crocce},
  {Castander}, {Fosalba}, {Marti}, {Miquel} \& {Cabr{\'e}}}{{Gazta{\~n}aga}
  et~al.}{2012}]{Gaztanaga12}
{Gazta{\~n}aga} E.,  {Eriksen} M.,  {Crocce} M.,  {Castander} F.~J.,  {Fosalba}
  P.,  {Marti} P.,  {Miquel} R.,    {Cabr{\'e}} A.,  2012, \mnras, 422, 2904

\bibitem[\protect\citeauthoryear{{Giavalisco}, {Ferguson}, {Koekemoer},
  {Dickinson}, {Alexander} \& {Bauer}}{{Giavalisco}
  et~al.}{2004}]{Giavalisco04}
{Giavalisco} M.,  {Ferguson} H.~C.,  {Koekemoer} A.~M.,  {Dickinson} M.,
  {Alexander} D.~M.,    {Bauer} F.~E. e.~a.,  2004, \apjl, 600, L93

\bibitem[\protect\citeauthoryear{{Ilbert}, {Capak}, {Salvato}, {Aussel},
  {McCracken}, {Sanders} \& {Scoville}}{{Ilbert} et~al.}{2009}]{Ilbert09}
{Ilbert} O.,  {Capak} P.,  {Salvato} M.,  {Aussel} H.,  {McCracken} H.~J.,
  {Sanders} D.~B.,    {Scoville} N. e.~a.,  2009, \apj, 690, 1236

\bibitem[\protect\citeauthoryear{{Joachimi} \& {Bridle}}{{Joachimi} \&
  {Bridle}}{2009}]{Joachimi09}
{Joachimi} B.,  {Bridle} S.~L.,  2009, ArXiv e-prints

\bibitem[\protect\citeauthoryear{{Jouvel}, {Kneib}, {Ilbert}, {Bernstein},
  {Arnouts}, {Dahlen}, {Ealet}, {Milliard}, {Aussel}, {Capak}, {Koekemoer}, {Le
  Brun}, {McCracken}, {Salvato} \& {Scoville}}{{Jouvel}
  et~al.}{2009}]{Jouvel09}
{Jouvel} S.,  {Kneib} J.-P.,  {Ilbert} O.,  {Bernstein} G.,  {Arnouts} S.,
  {Dahlen} T.,  {Ealet} A.,  {Milliard} B.,  {Aussel} H.,  {Capak} P.,
  {Koekemoer} A.,  {Le Brun} V.,  {McCracken} H.,  {Salvato} M.,    {Scoville}
  N.,  2009, \aap, 504, 359

\bibitem[\protect\citeauthoryear{{Kendall} \& {Stuart}}{{Kendall} \&
  {Stuart}}{1977}]{Kendall77}
{Kendall} M.,  {Stuart} A.,  1977, {The advanced theory of statistics. Vol.1:
  Distribution theory}

\bibitem[\protect\citeauthoryear{{Kennicutt}
  Jr.}{{Kennicutt}}{1998}]{Kennicutt98}
{Kennicutt} Jr. R.~C.,  1998, \araa, 36, 189

\bibitem[\protect\citeauthoryear{{Kirk}, {Lahav}, {Bridle}, {Jouvel}, {Abdalla}
  \& {Frieman}}{{Kirk} et~al.}{2013}]{kirkea_2013}
{Kirk} D.,  {Lahav} O.,  {Bridle} S.,  {Jouvel} S.,  {Abdalla} F.~B.,
  {Frieman} J.~A.,  2013, ArXiv e-prints

\bibitem[\protect\citeauthoryear{{Kirk}, {Laszlo}, {Bridle} \& {Bean}}{{Kirk}
  et~al.}{2011a}]{kirkea_2011}
{Kirk} D.,  {Laszlo} I.,  {Bridle} S.,    {Bean} R.,  2011a, ArXiv e-prints

\bibitem[\protect\citeauthoryear{{Kirk}, {Laszlo}, {Bridle} \& {Bean}}{{Kirk}
  et~al.}{2011b}]{MGPaper2}
{Kirk} D.,  {Laszlo} I.,  {Bridle} S.,    {Bean} R.,  2011b, ArXiv e-prints

\bibitem[\protect\citeauthoryear{{Le F{\`e}vre}, {Vettolani}, {Garilli},
  {Tresse}, {Bottini}, {Le Brun} \& {Maccagni}}{{Le F{\`e}vre}
  et~al.}{2005}]{LeFevre05}
{Le F{\`e}vre} O.,  {Vettolani} G.,  {Garilli} B.,  {Tresse} L.,  {Bottini} D.,
   {Le Brun} V.,    {Maccagni} D. e.~a.,  2005, \aap, 439, 845

\bibitem[\protect\citeauthoryear{{Martin}, {Fanson}, {Schiminovich},
  {Morrissey}, {Friedman}, {Barlow} T.~A.~{Conrow}, {Grange}, {Jelinsky} \&
  {Milliard}}{{Martin} et~al.}{2005}]{Martin05}
{Martin} D.~C.,  {Fanson} J.,  {Schiminovich} D.,  {Morrissey} P.,  {Friedman}
  P.~G.,  {Barlow} T.~A.~{Conrow} T.,  {Grange} R.,  {Jelinsky} P.~N.,
  {Milliard} B. e.~a.,  2005, \apjl, 619, L1

\bibitem[\protect\citeauthoryear{{McCall}, {Rybski} \& {Shields}}{{McCall}
  et~al.}{1985}]{McCall85}
{McCall} M.~L.,  {Rybski} P.~M.,    {Shields} G.~A.,  1985, \apjs, 57, 1

\bibitem[\protect\citeauthoryear{{Mouhcine}, {Lewis}, {Jones}, {Lamareille},
  {Maddox} \& {Contini}}{{Mouhcine} et~al.}{2005}]{Mouhcine05}
{Mouhcine} M.,  {Lewis} I.,  {Jones} B.,  {Lamareille} F.,  {Maddox} S.~J.,
  {Contini} T.,  2005, \mnras, 362, 1143

\bibitem[\protect\citeauthoryear{{Moustakas}, {Kennicutt} Jr. \&
  {Tremonti}}{{Moustakas} et~al.}{2006}]{Moustakas06}
{Moustakas} J.,  {Kennicutt} Jr. R.~C.,    {Tremonti} C.~A.,  2006, \apj, 642,
  775

\bibitem[\protect\citeauthoryear{{Padmanabhan}, {Budav{\'a}ri}, {Schlegel},
  {Bridges}, {Brinkmann}, {Cannon} \& {Connolly}}{{Padmanabhan}
  et~al.}{2005}]{Padmanabhan05}
{Padmanabhan} N.,  {Budav{\'a}ri} T.,  {Schlegel} D.~J.,  {Bridges} T.,
  {Brinkmann} J.,  {Cannon} R.,    {Connolly} A.~J. e.~a.,  2005, \mnras, 359,
  237

\bibitem[\protect\citeauthoryear{{Parkinson}, {Blake}, {Kunz}, {Bassett},
  {Nichol} \& {Glazebrook}}{{Parkinson} et~al.}{2007}]{Parkinson07}
{Parkinson} D.,  {Blake} C.,  {Kunz} M.,  {Bassett} B.~A.,  {Nichol} R.~C.,
  {Glazebrook} K.,  2007, \mnras, 377, 185

\bibitem[\protect\citeauthoryear{{Parkinson}, {Kunz}, {Liddle}, {Bassett},
  {Nichol} \& {Vardanyan}}{{Parkinson} et~al.}{2010}]{Parkinson10}
{Parkinson} D.,  {Kunz} M.,  {Liddle} A.~R.,  {Bassett} B.~A.,  {Nichol} R.~C.,
     {Vardanyan} M.,  2010, \mnras, 401, 2169

\bibitem[\protect\citeauthoryear{{Paykari} \& {Jaffe}}{{Paykari} \&
  {Jaffe}}{2013}]{Paykari13}
{Paykari} P.,  {Jaffe} A.~H.,  2013, \mnras, 433, 3523

\bibitem[\protect\citeauthoryear{{Polletta}, {Tajer}, {Maraschi}, {Trinchieri},
  {Lonsdale}, {Chiappetti} \& {Andreon}}{{Polletta} et~al.}{2007}]{Polletta07}
{Polletta} M.,  {Tajer} M.,  {Maraschi} L.,  {Trinchieri} G.,  {Lonsdale}
  C.~J.,  {Chiappetti} L.,    {Andreon} S. e.~a.,  2007, \apj, 663, 81

\bibitem[\protect\citeauthoryear{{Reid}, {Samushia}, {White}, {Percival},
  {Manera}, {Padmanabhan}, {Ross}, {S{\'a}nchez}, {Bailey}, {Bizyaev}, {Bolton}
  \& {Brewington}}{{Reid} et~al.}{2012}]{Reid12}
{Reid} B.~A.,  {Samushia} L.,  {White} M.,  {Percival} W.~J.,  {Manera} M.,
  {Padmanabhan} N.,  {Ross} A.~J.,  {S{\'a}nchez} A.~G.,  {Bailey} S.,
  {Bizyaev} D.,  {Bolton} A.~S.,    {Brewington} H. e.~a.,  2012, \mnras, 426,
  2719

\bibitem[\protect\citeauthoryear{{Schlegel}, {Abdalla}, {Abraham}, {Ahn},
  {Allende Prieto}, {Annis}, {Aubourg} \& {Azzaro}}{{Schlegel}
  et~al.}{2011}]{Schlegel11}
{Schlegel} D.,  {Abdalla} F.,  {Abraham} T.,  {Ahn} C.,  {Allende Prieto} C.,
  {Annis} J.,  {Aubourg} E.,    {Azzaro} M. e.~a.,  2011, ArXiv e-prints

\bibitem[\protect\citeauthoryear{{Tegmark}}{{Tegmark}}{1997}]{Tegmark97}
{Tegmark} M.,  1997, Physical Review Letters, 79, 3806

\bibitem[\protect\citeauthoryear{{Yamamoto}, {Parkinson}, {Hamana}, {Nichol} \&
  {Suto}}{{Yamamoto} et~al.}{2007}]{Yamamoto07}
{Yamamoto} K.,  {Parkinson} D.,  {Hamana} T.,  {Nichol} R.~C.,    {Suto} Y.,
  2007, \prd, 76, 023504

\bibitem[\protect\citeauthoryear{{Zoubian} \& {Kneib} J.-P.}{{Zoubian} \&
  {Kneib}}{2013}]{Zoubian13}
{Zoubian} J.,  {Kneib} J.-P. e.~a.,  2013

\end{thebibliography}

\section{Acknowledgements}
The authors thank the DESpec collaboration for their useful discussions which help developping this work.
Funding for this project was partially provided by the Spanish  
project AYA2009-13936, Consolider-Ingenio CSD2007- 00060, 
EC Marie Curie Initial Training Network CosmoComp (PITN-GA-2009-238356) 
and research project 2009- SGR-1398 from Generalitat de Catalunya. 
FBA thanks the Royal Society for support via an URF.

\end{document}